\documentclass[fleqn,10pt]{wlscirep}
\usepackage{booktabs}
\usepackage{lineno}
\usepackage{float}
\usepackage{hyperref}
\usepackage{bm}

\newcommand{\vc}[1]{\boldsymbol{#1}}

\title{Effective pruning of task-trained recurrent neural networks using noisy fluctuations and connection rescaling}

\author[1]{Sanjith Senthil}
\author[2,3,4*]{Rishidev Chaudhuri}
\affil[1]{The Harker School, San Jose, CA 95129, U.S.A.}
\affil[2]{Department of Neurobiology, Physiology and Behavior, University of California, Davis, Davis CA 95616, U.S.A.}
\affil[3]{Department of Mathematics, University of California, Davis, Davis CA 95616, U.S.A.}
\affil[4]{Center for Neuroscience, University of California, Davis, Davis, CA 95618, U.S.A.}
\affil[*]{rchaudhuri@ucdavis.edu}
\begin{document}

\begin{abstract}
Pruning network connections is key to brain function but, despite its importance, there exist few biologically-plausible pruning rules with demonstrated good performance. We evaluate noise-prune, a recently introduced unsupervised local pruning rule for recurrent networks that uses noisy fluctuations to determine the importance of connections. Noise-prune has previously only been empirically tested on random networks without a specific computational function. We show that noise-prune preserves task-performance in task-trained recurrent neural networks, greatly outperforming a strategy that only uses connection magnitude and exceeding a non-local strategy that uses second-order information. Rather than deterministically removing connections that fall below a threshold, noise-prune samples connections to preserve based on their importance and strengthens retained connections to preserve average strength. We show that this sampling and rescaling is essential to good performance, and can augment other pruning strategies. We also find that the optimal empirical degree of rescaling is lower than predicted by the original theoretical argument, and thus develop a new version of noise-prune that outperforms the previous version. Our results demonstrate that a biologically-motivated pruning strategy based on local activity statistics can effectively preserve task performance in functional recurrent networks by combining activity-dependent connection selection with compensatory strengthening of surviving connections.
\end{abstract}

\flushbottom
\maketitle
\thispagestyle{empty}

\section*{Introduction}
Pruning is the selective elimination of synaptic connections from a neural network, ideally while preserving its function. In biological neural networks, the selective elimination of synapses is crucial to development and plays an important functional role across the lifespan \cite{HuttenlocherDabholkar1997, kasai10, petanjek11, PaolicelliEtAl2011, stein19, yang09, lai12, Sakai2020}. Disruptions of synaptic elimination have been linked to multiple neurodevelopmental, psychiatric, and neurodegenerative disorders \cite{thomas2016over, SekarEtAl2016}. Synaptic transmission is energetically expensive and by removing synapses and maintaining sparse networks pruning likely also contributes to the brain’s impressive energy efficiency \cite{HarrisJolivetAttwell2012}, though the net energetic consequences of pruning also depend on neural activity and the strengths of retained synapses. Pruning is also of great interest in the development of efficient artificial neural networks \cite{lecun90, reed93, hassibi93, han15, dong17, bellec18, frankle19, narang17, lee19, baykal19, blalock20}, where it remains an area of active method development \cite{chengetal2024, frantarSparseGPT23}. 

The key question underlying a pruning rule is how to distinguish important from redundant connections. Connection strength is a useful proxy for importance, and magnitude-based pruning rules can be very successful in certain settings\cite{han15, zhu2017prune, gale2019sparsity}. However, large weights can be redundant and small weights can be important. The importance of a connection depends on the structure of the network as a whole, and changes over time for networks with dynamic input or recurrent connections. Thus, identifying the importance of a connection can be hard.

Most neural network pruning algorithms are developed for use in artificial neural networks \cite{lecun90, reed93, hassibi93, han15, dong17, bellec18, frankle19, narang17, lee19, baykal19, blalock20}. They are supervised, in that they have access to a cost function that measures task performance and they seek to remove connections while preserving task performance. They are moreover mostly developed for the setting of feedforward architectures, where information flows one way through a network and the downstream impact of a connection is comparatively easy to predict. This holds even for recent rules that take explicit inspiration from developmental pruning \cite{HanEtAl2024}. By contrast, biological neural networks likely do not have access to a measure of task performance during pruning (this is especially so if pruning happens during sleep) and synaptic plasticity rules must primarily rely on information local to the synapse \cite{GerstnerEtAl2018, LillicrapEtAl2020}. Biological neural networks are moreover typically highly recurrent, so that a connection’s impact depends on distributed time-varying loops of activity. Pruning under biological constraints is thus especially challenging. Consequently, despite the biological importance of pruning, the space of pruning rules that work under these biological constraints remains underexplored. 

This study analyzes noise-prune, a pruning rule previously introduced by Moore \& Chaudhuri\cite{moore2020noise} to model pruning in recurrent networks. Noise-prune seeks to preserve the dynamics of a recurrent neural network as connections are removed rather than directly optimizing task performance under pruning. Recurrent neural networks implement computations through their dynamics, and thus preservation of dynamics is a plausible task-independent strategy for maintaining computational function. Moreover, a biological network may support multiple computations and preserving dynamics should in principle preserve the network's response across these multiple computations rather than in a task specific way. 

The first key algorithmic idea behind noise-prune is that noisy fluctuations in neural activity encode information about the global dynamical structure of the network. If two neurons are much more correlated than one would expect given the strength of their direct connections, then they likely share information through higher-order loops or receive similar inputs, and the connection is likely to be redundant. Noise-prune uses the covariance of pairs of neurons in response to noise to locally estimate the importance of a connection and then stochastically either prunes or retains the connection. The second key idea is that retained connections should be strengthened to preserve the synapse's weight on average across pruning realizations, thus compensating for removed weights. 

Noise-prune thus combines two conceptually distinct operations: activity-dependent selection of connections to eliminate and compensatory strengthening of surviving connections. The connection-specific importance score and the compensatory strengthening depend only on synaptic strength and the activity statistics of the two neurons connected by the synapse, and not on an external cost function, thus satisfying an important local-information constraint on biological plasticity. Moreover, the two components of the rule have possible parallels with experimentally observed forms of synaptic plasticity. We return to these possible biological interpretations, and their limitations, in the Discussion.

In this study, we make three contributions. First, while previous work characterizing noise-prune’s performance only considered the preservation of dynamics (and not function) in random networks\cite{moore2020noise}, we apply noise-prune to recurrent neural networks trained to perform a set of tasks commonly used in neuroscience experiments \cite{yang2019, khona2023modcog}. We show that networks pruned with noise-prune preserve substantially more function than networks pruned purely on the basis of synaptic weight strengths. Noise-prune is moreover competitive with a more complex non-local pruning rule at low pruning fractions and outperforms it at high pruning fractions. Second, a notable feature of noise-prune is that connections are pruned stochastically, with a retention probability given by the noise-derived importance score, and connections that are not pruned are strengthened, rather than simply being preserved unchanged. We show that this sample-and-rescale strategy is essential to noise-prune's good performance. Moreover, such a sample-and-rescale strategy greatly improves the performance of a simpler pruning rule that only relies on the weight of a connection, thus suggesting a general compensatory principle. Finally, we characterize the optimal degree of rescaling of retained connections needed for good performance and show that it is lower than that predicted by the original theoretical argument. 

\section*{Results}
\begin{figure}[t]
\centering
\includegraphics[width=\linewidth]{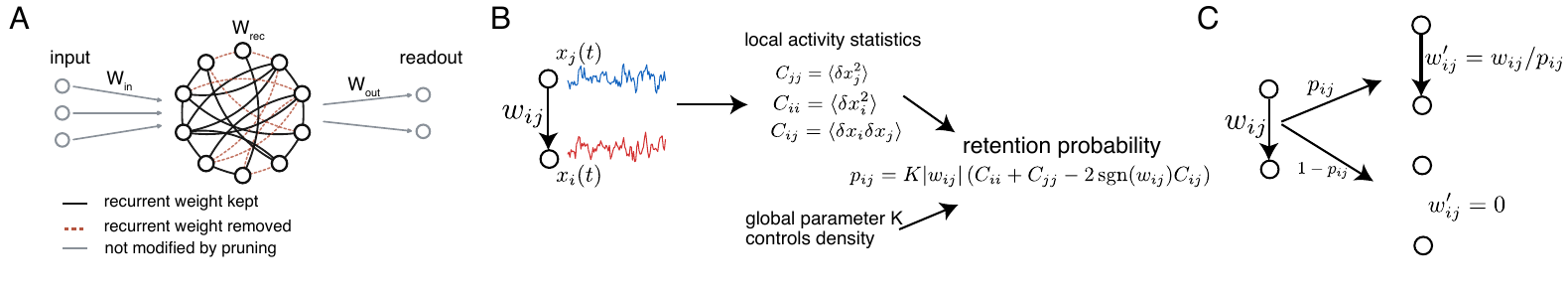}
\caption{\textbf{Schematic of network architecture and pruning algorithm.} (\textbf{A}) Schematic of recurrent network architecture. The network consists of an input, recurrent and output layer, with pruning taking place in the recurrent layer. (\textbf{B}) Computation of noise-prune's retention probability. Given a connection between a pair of neurons, the variances and covariance of the neurons are locally computed and combined with the connection weight and a global scaling factor to compute the retention probability. The retention probability is higher for stronger connections and, for positive weight connections, is lower if the two neurons are highly correlated. (\textbf{C}) Connections are either preserved, with probability given by the retention probability, or pruned. Preserved connections are strengthened by the inverse of the retention probability. Thus, the final strength of a retained connection depends inversely on the covariance term in panel B, and the weight is preserved in expectation across pruning realizations.}
\label{fig:schematic}
\end{figure}

\subsection*{The noise-prune algorithm}
We first briefly review the noise-prune algorithm\cite{moore2020noise}. Noise-prune is motivated by results from spectral graph sparsification, which seek to remove connections from a network while approximately preserving properties of its global structure and dynamics\cite{spielman11, spielman_teng11}. Specifically, sampling-based graph sparsification calculates a preservation probability for each edge in a graph and generates a sparser graph by sampling edges with these probabilities\cite{spielman11}. For an appropriate choice of probabilities, the spectrum of the graph Laplacian of the sparser graph is close to the spectrum of the original graph (and thus so are other properties of the graph that can be derived from the graph Laplacian). Moore \& Chaudhuri (2020) extended this framework to recurrent neural networks, observing that for restricted classes of linear and rectified-linear networks with symmetric, diagonally dominant connectivity, the sampling probabilities in sampling-based graph sparsification can be locally computed using correlations in noisy fluctuations in the network, allowing information about global network structure to be inferred from activity statistics available locally at each connection. These probabilities could then be used as part of a pruning rule that approximately preserved the spectrum of the network coupling matrix. Spectral preservation is relevant to recurrent neural networks because their computational properties depend on their dynamics, which are shaped by the spectrum of the recurrent connectivity. In linear networks, for example, the eigenvalues of the dynamical operator determine the timescales and oscillatory frequencies of different modes of activity. Preserving spectral properties therefore provides a principled, task-independent way to constrain aspects of network dynamics during pruning, which, in turn, may help preserve computational function. However, the theoretical guarantees do not extend to nonlinear asymmetric networks.

Consider pruning connections between neurons in an $N$-neuron recurrent neural network (Fig.~\ref{fig:schematic}). Let $x_i(t)$ be the activity of the $i$th neuron and $w_{ij}$ be the weight of the connection from neuron $j$ to neuron $i$. Let $C$ be the covariance matrix of network activity when the network receives noisy input.

For the connection from neuron $j$ to neuron $i$ define the retention probability
\begin{equation}\label{eq:pruning_rule}
p_{ij} =
\begin{cases}
Kw_{ij}\left(C_{ii} + C_{jj} - 2C_{ij}\right) & \text{for } w_{ij}>0 \quad \text{(excitatory)} \\
K|w_{ij}|\left(C_{ii} + C_{jj} + 2C_{ij}\right) & \text{for } w_{ij}<0 \quad \text{(inhibitory)}.
\end{cases}
\end{equation}
Here, the subscripts on $C$ denote entries in the covariance matrix, so that $C_{ij}$ is the covariance of the activity of neurons $i$ and $j$ and $C_{ii}$ and $C_{jj}$ are the variances of the $i$th and $j$th neurons respectively. The proportionality constant $K$ determines the overall density of the pruned network, with smaller $K$ corresponding to overall lower retention probabilities and thus sparser networks. Probabilities are capped at $1$.

For each connection, noise-prune independently chooses to preserve the connection with probability $p_{ij}$ and otherwise prunes it with probability $1 - p_{ij}$. Retained connections are strengthened by the rescaling factor $1/p_{ij}$, which maintains the connection strength in expectation. Thus, on average, noise-prune does not change the strength of a connection (but in any given realization of the rule many edges will be removed and others strengthened).

In the pruned network the $(i,j)$th weight is thus
\begin{equation}\label{eq:sampling_edges}
w^{sparse}_{ij} = 
\begin{cases}
w_{ij}/p_{ij} & \text{with probability } p_{ij} \\
0 & \text{otherwise}.
\end{cases}
\end{equation}

The retention probabilities depend on both the strength of a connection and the covariance of the activities of the connected neurons. The dependence on the strength is intuitive---all else being equal strong connections tend to be important. For positive (excitatory) connections, the covariance term can be written as \(C_{ii}+C_{jj}-2C_{ij}=\operatorname{Var}(x_i-x_j)\). Thus, connections between neurons whose activities fluctuate similarly receive lower retention probabilities, reflecting their potential redundancy with other network pathways. For negative (inhibitory) connections, the corresponding term is \(C_{ii}+C_{jj}+2C_{ij}=\operatorname{Var}(x_i+x_j)\), favoring connections between neurons whose activities fluctuate in opposite directions. Because these correlations depend on the full recurrent network, they incorporate information about its global structure while requiring only activity measurements from the two connected neurons (for more details and intuition see the original paper\cite{moore2020noise}).

The selection and strengthening components of noise-prune have an interesting relationship to neuronal activity covariance. For positive connections, holding synaptic strength and the individual neuronal variances fixed, increasing the covariance between the two neurons decreases the probability of retaining the connection. Thus, the connection-selection criterion is anti-Hebbian in character.

By contrast, the strengthening of surviving connections exhibits the opposite dependence. For a positive connection with retention probability below one, the post-pruning strength conditional on survival is
\[
w_{ij}^{\mathrm{sparse}}=\frac{w_{ij}}{p_{ij}}=\frac{1}{K(C_{ii}+C_{jj}-2C_{ij})}.
\]

This has two consequences. First, for fixed activity statistics, initially weaker connections receive greater proportional strengthening, and surviving connections approach the same final strength regardless of their initial weight. Second, holding the individual neuronal variances fixed, greater positive covariance leads to stronger connections conditional on survival. This latter dependence is Hebbian-like, in contrast to the anti-Hebbian character of the selection criterion.

Noise-prune thus uses correlations in two opposing ways: to identify potentially redundant connections for elimination, and to determine how strongly the surviving connections should be potentiated. The algorithm preferentially eliminates connections between neurons with similar fluctuations, while strengthening those connections more if they happen to survive. 

These relationships hold for the uncapped rule; limiting the rescaling factors, as we consider below, can attenuate them.

\begin{figure}[t]
\centering
\includegraphics[width=0.8\linewidth]{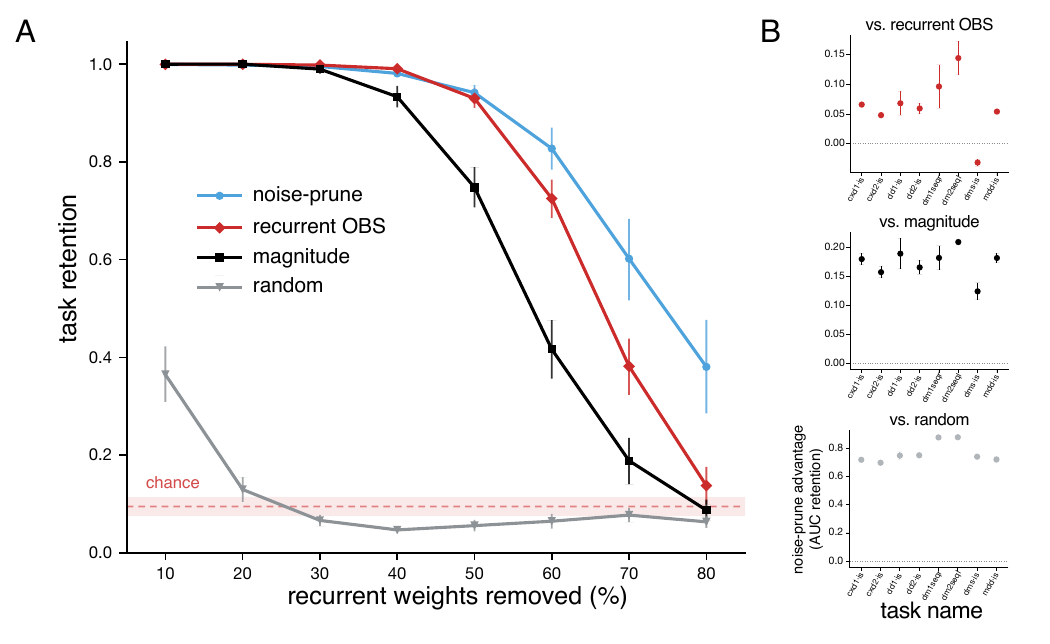}
\caption{\textbf{Task retention across pruning methods and sparsity levels.} (\textbf{A}) Task retention, the post-fixation sequence accuracy of each pruned network relative to its unpruned baseline, plotted against the percentage of recurrent weights removed, for noise-prune, recurrent OBS, magnitude pruning, and random pruning. Points show the mean over the $8$ tasks, and error bars show the $68\%$ bootstrap confidence interval. The dashed line and shaded band mark chance performance (the range of chance across tasks in units of retention). (\textbf{B}) The advantage of noise-prune over each other method on each task, measured as the difference in average retention across the pruning range (normalized area under the retention curve, $10\%$ to $80\%$). Points show the mean over the three networks for each task and error bars the $68\%$ bootstrap confidence interval across networks. Task names are abbreviated as follows: cxd1$\cdot$is, ctxdlydm1intseq; cxd2$\cdot$is, ctxdlydm2intseq; dd1$\cdot$is, dlydm1intseq; dd2$\cdot$is, dlydm2intseq; dms$\cdot$is, dmsintseq; mdd$\cdot$is, multidlydmintseq.}
\label{fig:task}
\end{figure}

\subsection*{Noise-prune preserves function in task-trained recurrent neural networks}
\label{results:pruning_performance}
Previous work on noise-prune has focused on proving theoretical guarantees and on empirically showing that the algorithm preserves the spectra and dynamics of random networks \cite{moore2020noise}, but noise-prune has not previously been tested on more realistic network architectures that perform a task. Task-trained recurrent neural networks are now common models of brain function \cite{yang2020artificial}. We thus sought to characterize noise-prune's performance on recurrent neural networks trained to perform a set of tasks commonly used in systems neuroscience and cognitive science. 

We draw these tasks from the recently introduced Mod-Cog battery of cognitive tasks \cite{khona2023modcog}, which extends a previous highly influential multi-task framework\cite{yang2019}. This newer battery of tasks was specifically designed to require non-trivial dynamical computations to solve and is thus especially appropriate for evaluating recurrent networks. 

We chose a set of $8$ tasks from the Mod-Cog battery, spanning working-memory, context-dependent stimulus selection, interval-estimation (\textit{int}), and sequence-production (\textit{seq}) subtasks, all of which require recurrent connectivity. Simpler decision-making variants (e.g., \textit{dm1}, \textit{anti}) either do not require recurrence or become trivially solvable as long as networks are not too small, making them uninformative for pruning comparisons (see Supplementary Fig.~S1). Task names and the performance of the unpruned networks are given in Supplementary Tables~S1 and~S2, and further details on the tasks and training procedure are provided in the Methods.

We then pruned these initially fully-connected task-trained networks to different levels of sparsity using noise-prune, in a one-shot fashion, and tested their performance without any retraining post-pruning. Note that multiple cycles of pruning and retraining would likely lead to higher performance but we consider a single cycle to isolate the effect of the pruning rule independent of the learning process. Also note that in our simulations pruning never improves overall task performance.

We compared noise-prune to three alternative pruning strategies. \emph{Random pruning} retains recurrent edges uniformly at random, subject to the fixed target density level, providing a structure-agnostic control. \emph{Magnitude pruning} ranks edges by their absolute weight, $s_{ij} = |(W_{\mathrm{rec}})_{ij}|$, and retains the top edges to match the target density. This baseline tests whether preserving the strongest recurrent couplings is sufficient to preserve network behavior. Magnitude pruning thus serves as a well-motivated and practical biologically-plausible comparison. Finally, we adapted the classic Optimal Brain Surgeon (OBS) algorithm \cite{lecun90, hassibi93} to preserve activity in a recurrent network, following the approach of a recent study that applied OBS layerwise to preserve activity in deep feedforward neural networks \cite{dong17}. This method, which we call \emph{recurrent OBS}, seeks to preserve post-pruning network activity rather than performance on a particular task, and is thus unsupervised. However, it is non-local and significantly more computationally-intensive than noise-prune. It relies on first inverting the Hessian matrix of the mean-squared difference between the pruned and unpruned neural activity (which can be related to the covariance matrix of neural activity) to estimate the sensitivity of network activity to changes in weights, and then uses this second-order information both to calculate the weights to prune and to compensate for the pruned weights by changing other weights in the network. It thus provides a strong theoretically-principled but more complex (and likely not biologically-plausible) comparison. 

We compared the pruning methods by task retention across pruning levels. The performance of noise-prune and these alternative strategies aggregated across the 8 tasks is shown in Fig.~\ref{fig:task}A.

To make quantitative comparisons between methods across sparsities, we analyze differences in the area under the retention curves of the two methods, with the sparsity range normalized to unit length. These differences directly capture the average improvement in task retention in the pruning range. 

We found that noise-prune significantly outperformed all $3$ alternative methods, with an average improvement in task retention across the pruning range of $0.174$ over magnitude pruning (95\% CI $[0.154, 0.193]$, $8/8$ tasks, $p = 0.0078$), $0.063$ over recurrent OBS (95\% CI $[0.029, 0.096]$, $7/8$ tasks, $p = 0.016$), and $0.769$ over random pruning (95\% CI $[0.726, 0.818]$, $8/8$ tasks, $p = 0.0078$). Random pruning rapidly collapses to near-chance retention ($< 0.110$) even at low sparsities, confirming that recurrent connectivity in these networks cannot be removed indiscriminately (see Methods for an explanation of slightly below chance performance at some sparsities). Notably, when evaluated across the full pruning range, noise-prune outperforms the computationally-intensive recurrent-OBS method. 

We also made two specific comparisons of noise-prune against the other methods at $60\%$ and $80\%$ weights pruned, highlighting noise-prune's gradual decline in task retention. At $60\%$ weights removed, noise-prune, OBS and magnitude-based pruning all performed well above chance, but noise-prune significantly outperformed the other three methods ($7/8$ tasks or better, $p \leq 0.031$ for all three comparisons). By $80\%$ weights removed, the other methods have collapsed to at or near-chance, but noise-prune still preserves a significant fraction of task performance and outperforms the other three methods ($8/8$ tasks, $p = 0.0078$ for all three comparisons). The full set of comparisons can be found in Supplementary Tables~S5 and S8.

The results so far aggregate performance across the 8 tasks but similar results hold on a single-task basis. In Fig.~\ref{fig:task}B, we show summary performance comparisons on a task-by-task basis (see Supplementary Fig.~S2 for retention curves for each of the $8$ tasks). Noise-prune outperforms magnitude and random pruning on all $8$ of the tasks, and outperforms recurrent OBS on $7/8$ tasks.
Thus, the better performance of noise-prune generalizes across tasks rather than being specific to a subset of tasks.

Our results thus show that noise-prune preserves performance in task-trained recurrent neural networks across tasks, and outperforms both magnitude-based pruning and the substantially more complicated recurrent-OBS strategy.  

\subsection*{Unpacking the contributions of the two components of noise-prune}
\begin{figure}[t]
\centering
\includegraphics[width=0.8\linewidth]{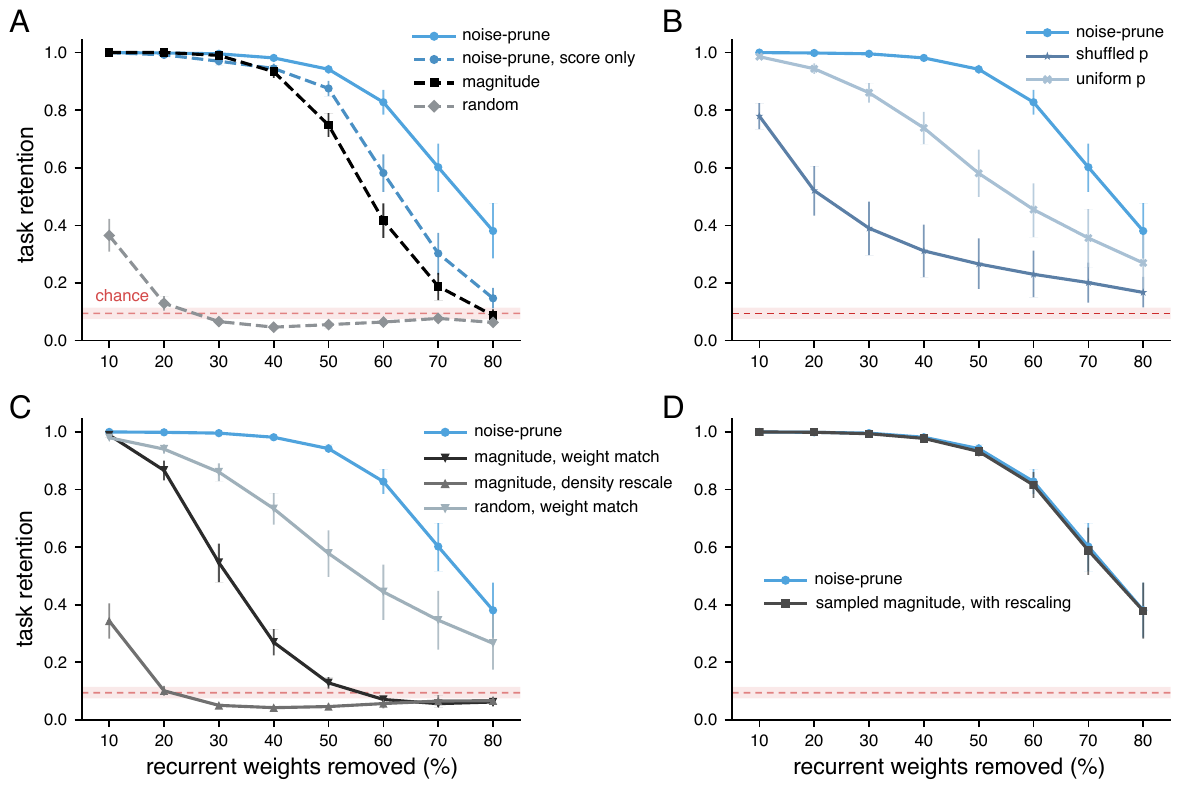}
\caption{\textbf{Task retention of noise-prune variants, magnitude variants, and rescaling controls across sparsity levels.} Task retention plotted against the percentage of recurrent weights removed. Points show the mean over the $8$ tasks and error bars the $68\%$ bootstrap confidence interval; the dashed line and shaded band mark chance. In every panel, noise-prune is the same trace as in Fig.~\ref{fig:task}. (\textbf{A}) Noise-prune compared with the deterministic variant that retains the top-scoring connections without sampling or rescaling (score only), magnitude pruning, and random pruning. (\textbf{B}) Sample-and-rescale with uninformative probabilities: every edge retained with probability equal to the target density (uniform $p$), or with the noise-prune probabilities randomly permuted across edges (shuffled $p$). (\textbf{C}) Deterministic pruning with a global rescaling of surviving weights: magnitude pruning with surviving weights rescaled to restore the total absolute recurrent weight (weight match) or by the inverse of the retained density (density rescale), and random pruning with surviving weights rescaled to restore the total absolute recurrent weight (note that random pruning with density rescale is almost identical to random weight match trace, so it is not plotted separately). (\textbf{D}) Sample-and-rescale with magnitude scores (sampled magnitude) compared to noise-prune.}
\label{fig:analyzing_rule}
\end{figure}

Noise-prune uses randomness in two ways, corresponding to the two key components of the pruning rule. First, noisy fluctuations are used to assign a covariance-based importance score to each connection. Second, these importance scores are used as part of a probabilistic sample-and-rescale process---importance scores are used as retention probabilities, and retained connections are strengthened by the inverse of the retention probability. Next, we sought to identify the specific contributions of each of these two components. 

We first removed the sample-and-rescale step and used the pruning probabilities derived by noise prune as deterministic importance scores. That is, we removed connections with the lowest importance scores to achieve the desired network sparsity. We found that removing the sample-and-rescale step significantly decreased performance below the original noise-prune, Fig.~\ref{fig:analyzing_rule}A (decrease in retention of $0.114$, 95\% CI $[0.095, 0.133]$, $8/8$ tasks, $p = 0.0078$). Nevertheless, this modified noise-prune rule with deterministic edge importance scores still significantly outperforms magnitude-based pruning (increase in retention of $0.060$, 95\% CI $[0.032, 0.086]$, $8/8$ tasks, $p = 0.0078$), showing that the noise-prune importance scores continue to have utility without the sampling and rescaling steps.

We also found that using noise to estimate the covariance matrix is crucial to achieving good pruning performance. Estimating the covariance matrix purely from task-driven input without noise significantly reduced retention by $0.107$ (95\% CI $[0.048, 0.188]$, $8/8$ tasks, $p = 0.0078$).

We next performed a complementary analysis, seeking to understand whether sampling-and-rescaling with uninformative scores would show good performance. We carried out sample-and-rescale using both uniform probabilities and shuffled noise-prune probabilities (thus preserving the marginal distribution of probabilities but not their assignment to particular edges). As shown in Fig.~\ref{fig:analyzing_rule}B, both strategies perform poorly (average task retention across the pruning range of $0.652$ for uniform probabilities, 95\% CI $[0.540, 0.791]$, and $0.342$ for shuffled probabilities, 95\% CI $[0.196, 0.531]$, compared to $0.862$ for noise-prune), showing that sampling-and-rescaling alone is inadequate for good performance.

The sample-and-rescale rule preserves the weight of a connection in expectation. A consequence is that it maintains expected recurrent input in the network. By contrast, pruning that simply removes edges without any further compensation causes overall recurrent network drive to decrease. We thus next asked whether the sample-and-rescale step simply acted to restore overall network gain. 

To test this gain preservation, we augmented magnitude-based pruning with a variety of homeostatic rescaling strategies that sought to preserve overall network drive. We first rescaled all retained connections by the inverse of the retained density (i.e., if $50\%$ of connections were removed then we doubled the strengths of the remaining connections). Second, we rescaled connections to preserve the sum of absolute weights (and thus total absolute recurrent weight). Both of these rescaling strategies performed very poorly (Fig.~\ref{fig:analyzing_rule}C; average retention of $0.081$, $95\%$ CI $[0.056, 0.103]$, and $0.352$, $95\%$ CI $[0.300, 0.395]$, respectively). Replacing noise-prune's per-edge probabilistic rescaling with the same two deterministic global factors was similarly ineffective (average retention of $0.092$, $95\%$ CI $[0.072, 0.109]$, and $0.302$, $95\%$ CI $[0.254, 0.345]$, respectively, traces not plotted). We then considered neuron-by-neuron rescaling strategies. To mimic standard synaptic rescaling \cite{turrigiano2008self}, we rescaled each neuron's incoming connections to match the pre-pruning total. This strategy also performed poorly (average retention $0.351$, $95\%$ CI $[0.293, 0.401]$, trace not plotted), as did the converse strategy of rescaling outgoing connections on a neuron-by-neuron basis (average retention $0.358$, $95\%$ CI $[0.299, 0.412]$, trace not plotted; Supplementary Table~S6). Thus, rescaling overall network drive does not explain the advantages of the sample-and-rescale strategy. Instead, probabilistic sampling of weights followed by strengthening retained weights on a synapse specific basis is crucial to good performance.

Given the success of the sample-and-rescale strategy, we next explored whether it could be used to enhance magnitude-based pruning. We constructed a hybrid pruning rule, which we call sampled-magnitude. Here we derive probabilities from appropriately normalized weight magnitudes, but we treat them as sampling probabilities and adopt the noise-prune strategy of rescaling each retained weight by the inverse of its sampling probability. We note that because this rescaling divides by weight, carrying out the rescaling sets all retained weights to the same magnitude except those with retention probabilities capped at one, thus serving to quantize the weights (see Discussion). We find that this hybrid approach dramatically enhances the performance of magnitude-based pruning (increase in retention of $0.168$, 95\% CI $[0.150, 0.186]$, $8/8$ tasks, $p = 0.0078$), placing it close to the performance of the original noise-prune rule (difference of $0.006$, 95\% CI $[-0.001, 0.013]$, $6/8$ tasks, $p = 0.12$). Thus, the sample-and-rescale strategy can be used as a general principle to augment existing pruning rules by converting deterministic importance scores into sampling probabilities.

\subsection*{Improving noise-prune with optimal rescale factors}
\begin{figure}[t]
\centering
\includegraphics[width=0.8\linewidth]{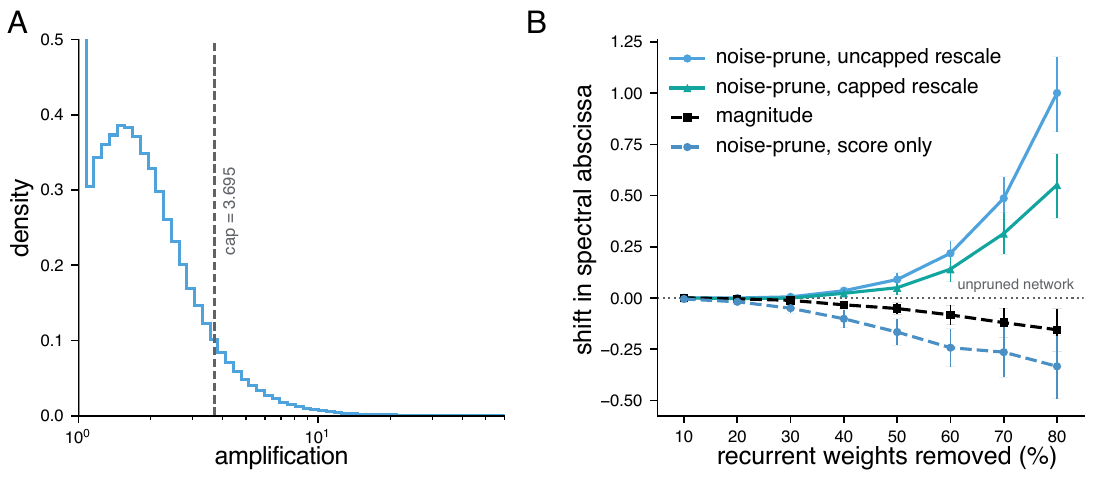}
\caption{\textbf{Amplification of surviving weights and spectral abscissa across pruning methods.} (\textbf{A}) Distribution of the amplification factors $1/p_{ij}$ applied to the weights that survive noise-prune at $70\%$ sparsity, pooled over all $24$ networks and three pruning seeds. The dashed line marks the constant cap $c = 3.695$ introduced below. Axes are truncated at a density of $0.5$ and an amplification of $60$ to show the bulk of the distribution, since it is heavy-tailed, with most surviving weights amplified by modest factors and a small number by factors in the hundreds or thousands. (\textbf{B}) Change in the spectral abscissa of the recurrent weight matrix from its unpruned value, plotted against the percentage of recurrent weights removed, for noise-prune with uncapped and with capped rescaling, magnitude pruning, and the score-only variant of noise-prune. Points show the mean over the $8$ tasks and error bars the $68\%$ bootstrap confidence interval.}
\label{fig:effect_of_uncapped_rescale}
\end{figure}

Together Figs.~\ref{fig:analyzing_rule}A and \ref{fig:analyzing_rule}D seem to present a puzzle. Fig.~\ref{fig:analyzing_rule}A shows that the noise-derived importance scores from noise-prune are substantially better at identifying connections to retain than simply using magnitudes as importance scores. By contrast, Fig.~\ref{fig:analyzing_rule}D suggests that this advantage is small when magnitude scores are used as sampling probabilities followed by rescaling---sampled magnitude performs nearly as well as noise-prune. Where does this discrepancy come from?

To resolve this puzzle, we first note that the rescaling factors depend inversely on the probabilities (as derived from a theoretical argument that seeks to preserve edge strength in expectation). As a consequence, edges sampled with very small probability receive very large multipliers. Figure~\ref{fig:effect_of_uncapped_rescale}A shows the distribution of the amplification factors applied to surviving weights at $70\%$ sparsity: it is skewed, with most edges amplified by modest factors (median $2.3$, mean $3.3$) but a small number amplified by very large factors (the largest factor applied in each network averaged $872$).

While biological synapses do show a heavy-tailed distribution of strengths \cite{buzsaki2014log}, these extreme per-edge amplifications are likely not biologically plausible. Moreover, such large connections could cause dynamical instability in a recurrent network. To probe the dynamical cost of these heavy-tailed multipliers, we computed the spectral abscissa for each pruned network. Noise-prune with rescaling raised the abscissa, increasingly with sparsity, by $1.00$ at $80\%$ sparsity (95\% CI $[0.61, 1.33]$) from an unpruned value of $1.09$, whereas magnitude pruning and the score-only variant of noise-prune, which do not rescale, lowered it (Fig.~\ref{fig:effect_of_uncapped_rescale}B). A larger abscissa corresponds to a potentially less stable dynamical regime, indicating a possible benefit to limiting amplification factors.

We reasoned that bounding this heavy-tail might improve the task performance of the pruned networks as well as increasing the biological plausibility of the rule. In particular, there should exist an optimal balance between rescaling to compensate for the lost connection weights and preserving stability. We thus introduced a capped rescale, which caps the amplification at the $q$-th percentile of the candidate-amplification distribution. 

\begin{figure}[t]
\centering
\includegraphics[width=0.8\linewidth]{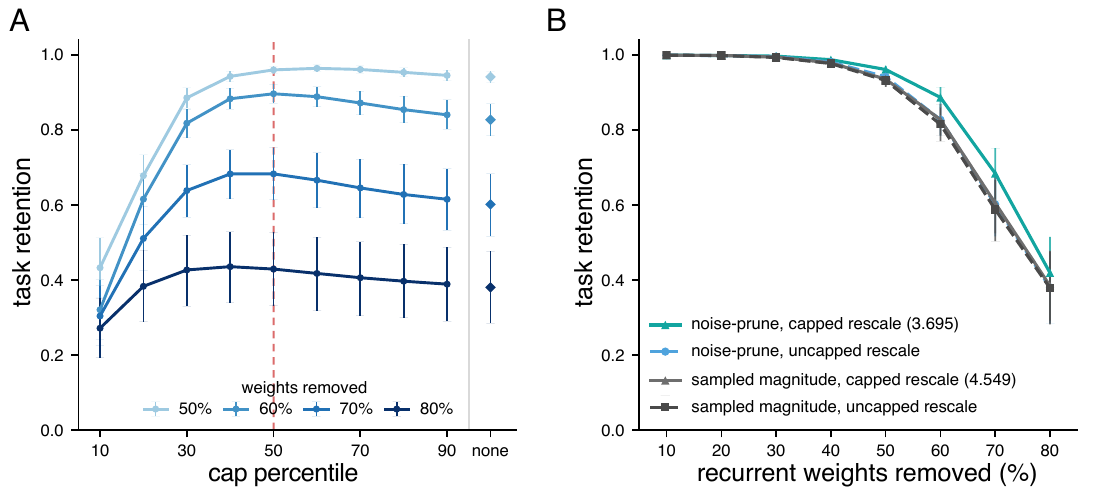}
\caption{\textbf{Task retention under capped rescaling across cap percentiles and sparsity levels.} (\textbf{A}) Task retention of noise-prune when the amplification of surviving weights is capped at the $q$th percentile of the candidate-edge amplification distribution, for $q = 10$ to $90$ at each sparsity level from $50\%$ to $80\%$, with the uncapped rule shown at right (none). The dashed line marks $q = 50$, the percentile at which retention pooled over the four sparsity levels peaks and from which the constant cap $c = 3.695$ is derived. (\textbf{B}) Task retention plotted against the percentage of recurrent weights removed for noise-prune and for sampled magnitude, each with its best constant cap ($3.695$ and $4.549$ respectively) and uncapped. Points show the mean over the $8$ tasks and error bars the $68\%$ bootstrap confidence interval.}
\label{fig:capped_task}
\end{figure}

Sweeping the cap strength for noise-prune reveals that task retention is non-monotonic in $q$. As shown in Fig.~\ref{fig:capped_task}A the smaller caps under-amplify and lose accuracy, the uncapped limit ($q = 100$ corresponding to original noise-prune) is suboptimal, and retention peaks at intermediate cap strengths ($q = 50$, though we note that performance does not change much across a range of cap strengths, and every cap from $q = 40$ to $q = 90$ outperformed the uncapped rule; Supplementary Table~S7 and also see Supplementary Fig.~S4 for the corresponding sweep for sampled magnitude). On average, the $50$th capping percentile corresponds to a rescale factor of $3.7$.

Motivated by these observations, we introduced a new variant of noise-prune with the rescale factors capped at a constant factor of $3.7$ (rather than choosing the cap factor on a run-by-run basis, which seemed less plausible). The capped noise-prune rule increased the average task retention by $0.027$ when compared to the original rule (95\% CI $[0.014, 0.040]$, $7/8$ tasks, $p=0.016$) and limited the rise in spectral abscissa at $80\%$ sparsity from $1.00$ to $0.55$ (95\% CI $[0.22, 0.84]$; Fig.~\ref{fig:effect_of_uncapped_rescale}B). Thus, while rescaling is critical for good performance, optimal performance corresponds to modest rescaling.

We compared this improved version with the other pruning rules. We also applied the same capping procedure to the sampled magnitude rule, and chose its best performing cap for inclusion. However, the magnitude rule only very slightly benefited from capping (average task retention increased by $0.0055$, 95\% CI $[0.0021, 0.0087]$, $6/8$ tasks, $p = 0.031$). We found that the new variant of noise-prune outperformed all the previous pruning rules. In particular, it outperforms both capped sampled magnitude (increase in retention of $0.027$, 95\% CI $[0.015, 0.039]$, $8/8$ tasks, $p=0.0078$) and recurrent OBS (increase in retention of $0.090$, 95\% CI $[0.065, 0.115]$, $8/8$ tasks, $p=0.0078$). Thus, we find that capping the rescale factors leads to a better version of noise-prune.

\section*{Discussion}
The pruning of connections in the brain operates under challenging biological constraints: networks are highly-recurrent, making the importance of each synapse non-stationary and hard to estimate, and plasticity rules must rely on the comparatively limited information available locally at a synapse. In this study, we examined the performance of noise-prune, a previously proposed biologically-plausible pruning rule that uses noisy fluctuations in neural activity to probe higher-order network structure and prune redundant connections while strengthening retained connections \cite{moore2020noise}. We showed that noise-prune preserves performance in recurrent neural networks trained to perform cognitive tasks and outperforms several competing strategies. We unpacked the relative contributions of the two major components of noise-prune and, moreover, showed that sampling connections and rescaling by the inverse sampling probability is also a useful strategy for magnitude-based pruning. Finally, we found that the optimal degree of rescaling is lower than that predicted by theory and that using this optimal rescaling leads to an improved version of noise-prune.

\subsection*{General computational principles underlying noise-prune}
The noise-prune algorithm was originally motivated by very strong theoretical results from sampling-based spectral graph sparsification\cite{spielman11, spielman_teng11}. Building on these results, for linear and rectified-linear networks with symmetric diagonally-dominant connectivity, noise-prune preserves the entire spectrum of the network operator to within a multiplicative factor \(\epsilon\), using only \(O(N\log N/\epsilon^2)\) connections \cite{spielman11, moore2020noise}. Thus, as network size increases, the entire spectrum can be approximately preserved even as the fraction of retained connections approaches zero. This spectral preservation in turn preserves network dynamics. These theoretical guarantees significantly exceed what is needed for biological pruning, where connection density is not expected to asymptotically approach $0$ and where networks show redundancy and low-dimensional structure. 

However, the theoretical guarantees do not apply to the nonlinear asymmetric networks studied here, nor do they establish a general relationship between spectral preservation and task-dependent computational function. Thus, although preserving network dynamics provides a plausible task-independent strategy for preserving computational function, whether noise-prune preserves task performance in general task-trained recurrent networks is an empirical question that we addressed in this study. Our results empirically demonstrate that noise-prune can preserve task performance in networks beyond the theoretically established setting. The good performance of noise-prune on a wider class of task-trained recurrent neural networks suggests that it may be possible to provide weaker but more general theoretical guarantees on the performance of noise-prune and on the formal relationship between dynamical preservation and task performance in recurrent networks.

\subsection*{The role of probabilistic sampling and rescaling}
Noise-prune uses noisy fluctuations to estimate an importance score for each connection but then, instead of pruning connections whose scores fall below a certain threshold, it uses these scores as sampling probabilities. Connections are retained with probability proportional to their score, and retained connections are strengthened by the inverse of the sampling probability, yielding a final weight that depends on the covariance of the two connected neurons but not on their initial weight. This sample-and-rescale strategy allows the pruning rule to appropriately compensate for pruned edges. For example, consider a set of connections that are redundant with each other but that together play an important role in a network’s function so that it is important to preserve at least a few of these connections. Noise-prune would assign all these edges a low importance score but the probabilistic sampling would allow some of these edges to survive and be greatly strengthened, thus effectively replacing the set by a few representative members that perform the same function. By contrast, a deterministic pruning rule that used a strict threshold would either need to assign these connections importance scores above the threshold to ensure that some are preserved or would need to recalculate importance scores after each edge is pruned (like the OBS-derived approaches) rather than carrying out one-shot pruning. By comparing noise-prune with and without probabilistic sampling and rescaling we empirically showed that the sampling approach is crucial for good performance---deterministic variants perform much worse.

While rescaling is essential, we found that capping the degree of rescaling by upper bounding the rescale factors leads to improved task performance, likely as a consequence of maintaining network stability. Thus, in a more realistic recurrent network, the optimal degree of rescaling is lower than that predicted by theory. We note that such a cap naturally corresponds to a cap on either how much a synapse can grow or on the maximum size of a synapse and thus is easy to implement. Moreover, improved performance is seen for a wide range of caps and thus a specific value of the cap is not crucial.

Interestingly, we found that the sample-and-rescale approach greatly improved the performance of magnitude-based pruning. This benefit was not seen with other forms of gain compensation, including overall scaling of connections in the network and compensations that, motivated by homeostatic synaptic scaling \cite{turrigiano2008self}, preserved the total input or output of each individual neuron. Thus, sample-and-rescale could potentially benefit a wider variety of pruning rules. The magnitude-based case is particularly interesting because, for retention probabilities $p_{ij}=K|w_{ij}|$, rescaling retained weights by $1/p_{ij}$ yields weights of $\pm 1/K$, independent of their original magnitudes. Consequently, apart from connections whose retention probabilities are capped at one, the resulting network has only three possible weight values: $-1/K$, $0$, and $+1/K$. This is reminiscent of ternary-weight quantization in artificial neural networks, where restricting weights to a small set of discrete values can reduce memory and computational costs \cite{han2016deep,zhu2017trained}. Our results thus suggest that stochastic pruning with rescaling may provide a complementary route to obtaining sparse, low-precision networks.

\subsection*{Alternative approaches to pruning}
To provide a sophisticated and well-motivated comparison to noise-prune, we slightly adapted the layer-wise Optimal Brain Surgeon algorithm\cite{hassibi93, dong17} for use in a recurrent neural network (shown in results under the name recurrent-OBS). We found that noise-prune performed on par with recurrent-OBS for sparsities up to about 50\% and outperformed it at higher sparsities. This better performance of noise-prune is surprising, given that recurrent-OBS is a more complex and non-local method.  In principle, applying OBS requires reevaluating the Hessian of the loss function after pruning each weight. However, to provide an equivalent comparison to the one-shot pruning of noise-prune (and for computational tractability) we also applied this OBS variant in a one-shot fashion and did not reevaluate the Hessian after each pruning. It is possible that recurrent-OBS would perform better if allowed multiple cycles of Hessian evaluation and pruning, and comparing pruning rules in this multiple cycle regime is a natural direction for future work. 

An elegant complementary perspective on synaptic pruning comes from Bayesian model reduction and statistical model selection, in which network connectivity is optimized by balancing the ability to explain observations against model complexity\cite{mackay1992practical, friston2011post, markovic2024bayesian, tazawa2025synaptic}. For example, Markovi\'c et al. (2024) develop a stochastic Bayesian model-reduction approach that uses approximate posterior distributions over network weights and efficiently estimates the evidence for removing connections, iteratively combining pruning with further network optimization. More closely related to synaptic plasticity, Tazawa and Isomura (2026) derive a biologically-plausible online pruning rule that implements Bayesian model selection within a generative-model framework, with connectivity variables updated through activity-dependent mechanisms. These approaches provide principled methods for identifying sparse network structures and can improve inference and learning. Noise-prune addresses a complementary problem: sparsifying an existing recurrent network while seeking to preserve its dynamical properties, without requiring a probabilistic generative model, posterior inference over synaptic weights, or access to task performance during pruning. Instead, it uses fluctuations in neural activity to estimate the dynamical importance of individual connections and probabilistically samples and strengthens retained connections. This provides a relatively simple approach to sparsification using connection-specific information available locally to synapses (although a global parameter is required to control overall connection density). Unlike Bayesian model-selection approaches, noise-prune does not explicitly optimize model evidence or the trade-off between predictive performance and complexity. Moreover, its theoretical guarantees apply to restricted classes of networks, specifically linear and rectified-linear symmetric diagonally-dominant networks.

\subsection*{Biological interpretation and plausibility}
The biological motivation for noise-prune can be understood in terms of two conceptually distinct processes: an activity-dependent mechanism for selecting the connections to eliminate along with a compensatory mechanism for strengthening surviving connections. The algorithm couples these operations mathematically, but they likely correspond to distinct biological plasticity mechanisms.

First, activity can influence synapse elimination during developmental circuit refinement. For example, microglia preferentially engulf relatively less active retinal inputs in the developing retinogeniculate system\cite{schafer2012microglia}. The activity-dependent selection criterion in noise-prune has an anti-Hebbian character for positive connections: greater covariance between pre- and postsynaptic neurons reduces the probability of retaining the connection. Thus, it favors the elimination of connections between neurons with similar fluctuations, which may be redundant due to alternative pathways within the recurrent network. Anti-Hebbian forms of plasticity have been experimentally observed in several neural circuits, including the electrosensory system of weakly electric fish\cite{bell1997synaptic}, the mammalian dorsal cochlear nucleus\cite{tzounopoulos2004cell, tzounopoulos2007coactivation}, and excitatory connections between cortical pyramidal neurons under particular conditions\cite{letzkus2006learning}. These findings show that correlated activity can promote synaptic depression rather than potentiation, and anti-Hebbian mechanisms have been implicated in suppressing predictable sensory signals. However, these studies are concerned with changes in synaptic strength, typically induced by specific spike-timing relationships, rather than structural elimination of connections according to overall covariance. Thus, the anti-Hebbian interpretation of noise-prune provides a possible biological analogy but is not an established mechanism of synapse elimination.

A separate component of noise-prune is the strengthening of surviving connections. Compensatory strengthening accompanies synapse elimination during biological circuit refinement; for example, developmental elimination of retinal inputs to thalamic neurons is accompanied by substantial strengthening of the remaining connections\cite{chen2000developmental}. Homeostatic synaptic plasticity can also produce approximately multiplicative changes in synaptic efficacy\cite{turrigiano1998activity}, including compensatory modifications specific to individual synapses\cite{beique2011arc}. Capping rescaling as we propose increases the biological plausibility of the rule and its connection to these mechanisms. The uncapped rescale factors can run into the thousands, an unrealistic factor even accounting for the heavy-tailed distribution of synaptic strengths and likely out of range of such homeostatic mechanisms. By contrast, we find optimal performance when rescaling is capped at relatively modest and biologically-plausible factors of $3$ to $6$ (Supplementary Table~S7).

Interestingly, the compensatory component of noise-prune exhibits a Hebbian-like dependence on activity covariance, in contrast to the anti-Hebbian character of the selection criterion. For positive connections, the strength of a surviving synapse is inversely proportional to the variance of the difference between presynaptic and postsynaptic activity (and thus, expanding the inverse to first-order, it is proportional to covariance). This produces two qualitative predictions: initially weaker synapses undergo greater proportional strengthening if they survive, and connections between more strongly correlated neurons become stronger if they survive (holding other relevant quantities fixed). The first prediction is consistent with observations that small dendritic spines are particularly susceptible to persistent potentiation\cite{matsuzaki2004structural} and that they tend to either vanish or to strengthen and stabilize \cite{holtmaat05}, while the second is consistent with the general Hebbian principle that coordinated pre- and postsynaptic activity can induce synaptic strengthening. These observations provide possible biological support for the strengthening mechanism.

The factor by which surviving connections are strengthened in noise-prune is derived mathematically to preserve synaptic weights in expectation, rather than from a specific biological mechanism. Nevertheless, for positive connections with retention probabilities below one, the final strength of a surviving connection is independent of its original strength and depends only on a statistic of local activity. This suggests a possible interpretation in which surviving synapses approach activity-dependent target strengths through local plasticity mechanisms, rather than explicitly calculating inverse retention probabilities. We also note that our results show that limiting the degree of potentiation improves task performance, suggesting that approximate inverse-probability compensation is sufficient and that biologically realistic limits on synaptic potentiation may be compatible with, and potentially beneficial for, the compensatory mechanism.

\subsection*{Limitations}
Our study has several limitations. First, the networks we consider use firing-rate dynamics and do not obey Dale's law, allowing individual neurons to have both excitatory and inhibitory outgoing connections. Whether noise-prune performs similarly in more biologically realistic networks, including spiking networks with separate excitatory and inhibitory populations, remains to be established. Second, although noise-prune uses local synaptic information to determine which connections to retain, it requires a global scaling factor ($K$) to control overall connection density. Assuming that the amplitude of noisy fluctuations is set by ongoing biological processes, this leaves a single network-wide parameter that must be regulated for pruning, along with potentially a second parameter that limits the maximum strengthening of retained connections or the maximum size of synapses. How these global parameters might be set or regulated biologically remains an open question. Finally, we evaluate noise-prune on a limited set of cognitive tasks, with each network trained on only a single task. An important direction for future work is to test noise-prune in networks trained to perform multiple tasks, a setting that was a central motivation for the original algorithm.

In sum, our results demonstrate that task performance in recurrent neural networks can be preserved during substantial pruning by combining activity-dependent estimates of connection importance with probabilistic synapse selection and compensatory strengthening. The rule is motivated by local-information constraints and has parallels with experimentally observed forms of synaptic plasticity, but its precise biological implementation remains an open question.

\section*{Methods}
\subsection*{Network architecture}
We model recurrent neural circuitry using continuous-time recurrent neural networks (CTRNNs) with firing-rate dynamics. Let $v(t) \in \mathbb{R}^{H}$ be the hidden-state voltage at time $t$, $r(t) \in \mathbb{R}^{H}$ the corresponding firing-rate state, and $u(t) \in \mathbb{R}^{I}$ the input, where $H$ is the number of hidden units and $I$ the input dimension. The continuous-time dynamics are
\begin{equation}
\tau \dot{v}(t) = -v(t) + W_{\mathrm{in}} u(t) + b_{\mathrm{in}}
+ W_{\mathrm{rec}} r(t) + b_{\mathrm{rec}},
\label{eq:ctrnn}
\end{equation}
with firing rates
\begin{equation}
r(t) = \phi\!\left(v(t)\right),
\label{eq:rates}
\end{equation}
where $W_{\mathrm{in}} \in \mathbb{R}^{H \times I}$ is the input weight matrix, $W_{\mathrm{rec}} \in \mathbb{R}^{H \times H}$ the recurrent weight matrix, $\tau$ the scalar membrane time constant, and $b_{\mathrm{in}}, b_{\mathrm{rec}} \in \mathbb{R}^{H}$ the input and recurrent biases. The activation function $\phi(v) = \tanh(v)$ is used throughout this study. The network readout is
\begin{equation}
y(t) = W_{\mathrm{out}} r(t) + b_{\mathrm{out}},
\label{eq:readout}
\end{equation}
with $W_{\mathrm{out}} \in \mathbb{R}^{O \times H}$ and $b_{\mathrm{out}} \in \mathbb{R}^{O}$ mapping firing rates to an $O$-dimensional task output, where $O$ is the output dimension. The CTRNN was simulated using a forward Euler discretization. Defining
$\alpha = dt/\tau$, the discrete update is
\begin{equation}
v_t = v_{t-1} + \alpha\!\left(-v_{t-1} + W_{\mathrm{in}} u_t
+ b_{\mathrm{in}} + W_{\mathrm{rec}} r_{t-1} + b_{\mathrm{rec}}\right),
\label{eq:euler}
\end{equation}
where the nonlinear rate update is $r_t = \phi(v_t)$.

\subsection*{Network suite}
We trained a suite of CTRNNs ($8$ tasks $\times$ $3$ network seeds) with hidden size $H = 512$ on eight Mod-Cog cognitive tasks\cite{khona2023modcog,yang2019}. In each task the network receives ring-encoded stimuli and produces a ring-encoded sequential response over a trial, structured into fixation, stimulus, and (where present) delay periods. The selected variants span working-memory delays, context-dependent stimulus selection, interval-estimation (\textit{int}), and sequence-production (\textit{seq}) subtasks, which require recurrent connectivity. 

Simpler decision-making variants (e.g., \textit{dm1}, \textit{anti}) either do not require recurrence or become trivially solvable at $H = 512$, making them uninformative for pruning comparisons (see Supplementary Fig.~S1). The eight task names are given in Supplementary Table~S2, and their formal specifications are provided in Supplementary Table~S1. For each task we trained three networks with independent weight initializations, yielding $24$ networks total. To test whether our results depend on network size, we additionally trained the same suite of 24 networks with $H = 1024$ and report pruning results for these networks in Supplementary Fig.~S5.

All networks use the architecture and discretization described above, with $dt = 10$ ms and membrane time constant $\tau = 100$ ms, giving $\alpha = dt/\tau = 0.1$. Self-connections were not allowed throughout training and testing in all experiments, so the diagonal of $W_{\mathrm{rec}}$ was set to zero. The total number of prunable recurrent edges is therefore $H(H-1) = 261{,}632$ per network for $H=512$, and all pruning targets are specified as fractions of these off-diagonal edges.

Networks were trained with Adam\cite{kingma2017adam} (default $\beta_1 = 0.9$, $\beta_2 = 0.999$, $\epsilon = 10^{-8}$) and batch size 256. Targets $Y \in \mathbb{Z}^{T \times B}$, over $T$ timesteps and $B$ batch elements, encode invalid positions as $-1$. Networks were evaluated on task outputs and not their ability to maintain fixation. No weight decay or recurrent-weight penalty was applied, so $W_{\mathrm{rec}}$ develops without regularization bias.

Training proceeded in two phases: an initial 6,000-step phase at learning rate $1.2\times10^{-3}$, followed by a continuation phase of 6,000 additional steps at learning rate $6\times10^{-4}$ (12,000 steps total), with gradient norm clipped at 1.0 throughout. 

Networks were evaluated every 500 training steps on 64 validation batches. The final checkpoint for each network was selected as the training step that maximized validation sequence accuracy, using cross-entropy loss as a tie-break. The performance of the trained networks on each task is summarized in Supplementary Table~S2 (and is on par with results reported in the original Mod-Cog paper\cite{khona2023modcog}).

\subsection*{Pruning methods}
We considered three ways of scoring the importance of a connection and two ways of pruning based on the scores. Connections were scored using the noise-derived covariance score of noise-prune, the absolute weight of the connection (magnitude), or a uniform score (random). Given the scores, we either pruned deterministically, retaining the top-scoring connections to match a target density, or followed noise-prune's sample-and-rescale approach, in which connections are retained with probability given by the score and retained connections are strengthened by the inverse of that probability, with or without a cap on the rescaling. We additionally consider a set of controls in which sample-and-rescale is applied with uninformative probabilities or in which pruned networks are rescaled by a single global factor. We also compare against a much more sophisticated non-local method based on the Optimal Brain Surgeon (OBS) algorithm \cite{lecun90, hassibi93, dong17}. All pruning was applied only to $W_{\mathrm{rec}}$, leaving input and readout weights untouched, and diagonal entries of the matrix (i.e., self-connections) are maintained at $0$.

Random pruning retains recurrent edges uniformly at random, subject to the fixed target density level, providing a structure-agnostic control. Magnitude pruning ranks edges by their absolute weight, $|w_{ij}|$, and retains the top edges to match the target density. This is a well-established biologically-plausible baseline method and it tests whether strong raw recurrent couplings are sufficient to preserve network behavior. Thus, it serves as a strong practical comparison.

We describe noise-prune, the sample-and-rescale variants and controls, the capped rescaling, and OBS below.

\subsubsection*{Noise-prune}
Noise-prune calculates the retention probability $p_{ij}$ for each edge from the weight $w_{ij}$ and the covariance matrix $C$ of network activity in response to noise, as in Eq.~\eqref{eq:pruning_rule}, with probabilities exceeding 1 clipped to $1$ and $K$ chosen so that the expected number of retained connections matches the target density. In the case of linear diagonally-dominant networks of size $N$ with unit variance noise, $K$ needs to be at least $8\log(N)/\epsilon^2$ to preserve the spectrum to within a multiplicative factor of $\epsilon$ (see Moore \& Chaudhuri \cite{moore2020noise} for further details). Each connection is then independently retained with probability $p_{ij}$ and, if retained, strengthened by $1/p_{ij}$ (Eq.~\eqref{eq:sampling_edges}).

Noise-prune has previously been applied only to linear and rectified-linear networks, with the covariance matrix either estimated by simulating the network in response to noise or exactly calculated using the Lyapunov equation for a noisy linear system \cite{moore2020noise}. Since a linear approximation to the covariance matrix is poor given the significant nonlinearity of the trained networks, here we estimate an empirical covariance matrix $\hat{C}$ from noise-driven simulation of the nonlinear network. Noise is added to the firing rates at each Euler step of the discrete network dynamics. 
\begin{equation}
r_t \leftarrow \tanh(v_t) + \sigma\,\xi_t, \qquad \xi_t \sim \mathcal{N}(0, I).
\label{eq:noise_injection}
\end{equation}

In a linear network, scaling the noise standard deviation $\sigma$ simply rescales the covariance matrix and hence rescales the probabilities. Thus, $\sigma$ can be absorbed into the overall proportionality constant $K$. For a nonlinear network, however, different noise magnitudes lead to the network exploring different volumes of state space and produce different covariances and hence probabilities. We thus set $\sigma$ to match the network's natural voltage variability under noise-free task input. Letting $v^{\mathrm{task}}$ denote the voltage trajectory under task input with no injected noise, we set $\sigma = \sigma_{\mathrm{nat}}$ with
\begin{equation}
\sigma_{\mathrm{nat}} = \sqrt{\frac{1}{H}\sum_{i=1}^{H}\operatorname{Var}\!\big(v_i^{\mathrm{task}}\big)}.
\label{eq:sigma_nat}
\end{equation}
We further verified that the set of connections that received the top $50\%$ of sampling probabilities were robust to changes in $\sigma$ (set overlaps by $>95\%$ as $\sigma$ is changed from $0.75$ to $1.5\times$ $\sigma_{\mathrm{nat}}$).

To ensure that the network explores the appropriate regions of state space, trajectories are simulated under task-input batches (from the task on which the network was trained) with injected noise. From each rollout we take the final-timestep states across its batch of trials, accumulating over repeated independently-noised rollouts until $M = 25{,}000$ samples were collected. We center each rollout separately before pooling: for rollout $c$ with samples $\{r_{t,b}\}_{(t,b)\in c}$, we subtract that rollout's mean, $\delta r_{t,b} = r_{t,b} - \bar{r}_c$ with $\bar{r}_c = \frac{1}{M_c}\sum_{(t,b)\in c} r_{t,b}$, and concatenate the centered samples across rollouts ($M = \sum_c M_c$). The empirical covariance is then
\begin{equation}
\hat{C} = \frac{1}{M-1} \sum_{c}\sum_{(t,b)\in c}
\delta r_{t,b}\,\delta r_{t,b}^{\top}.
\label{eq:emp-cov}
\end{equation}
Conditional centering, which subtracts a noise-free reference trajectory from the current trajectory at each time-step, isolates noise-induced perturbations more cleanly, but is computationally more expensive since it requires running network simulation twice (once baseline, once noisy). We evaluated conditional centering as an ablation and found very similar task-preservation results, indicating that centering choice has a small effect. Trajectory-mean centering is thus used throughout for its computational feasibility and similar performance, and because it is more biologically plausible, possibly representing a slow moving average of activity.

We also evaluated three variants of the covariance estimate: one in which the network receives injected noise only, without task input, one in which states are sampled from all timesteps of each rollout rather than the final timestep, and one combining both. All three yielded very similar task retention to the estimate above (Supplementary Table~S4), indicating that the scores reflect the network's response to injected noise rather than task-driven activity or the sampling window. By contrast, estimating the covariance from task input with no injected noise reduced retention averaged over the pruning range from $0.862$ to $0.756$ (difference $0.107$, 95\% CI $[0.048, 0.188]$, $8/8$ tasks, $p = 0.0078$), so the injected noise is essential to the rule.

As noise-prune is probabilistic, the desired density is only guaranteed in expectation and post-pruning density may differ slightly from the desired density (these fluctuations are proportionally smaller as the network gets bigger). However, given both the large network sizes and the skewed distribution of preservation probabilities, in our simulations noise-prune almost exactly matches the target density. In any given realization, the fluctuations are always smaller than $0.2$ percentage points of sparsity ($0.05$ percentage points on average).

For the comparison in Fig.~\ref{fig:analyzing_rule}, we also consider a variant of noise-prune without sampling and rescaling. To implement this variant we compute the same scores but rather than treating them as probabilities and first sampling and then rescaling preserved edges, we deterministically rank edges by their scores and retain the top fraction matching a target density, without altering surviving weights.

\subsubsection*{Sample-and-rescale with magnitude scores}
For the sampled-magnitude rule, we used the absolute weight $|w_{ij}|$ in place of the covariance-based score in Eq.~\eqref{eq:pruning_rule}, so that $p_{ij} = K|w_{ij}|$, with $K$ chosen as for noise-prune so that the expected number of retained connections matches the target density. We then sampled and rescaled connections exactly as in noise-prune (Eq.~\eqref{eq:sampling_edges}), rather than deterministically pruning. The capped rescaling described below was applied to this rule in the same way as to noise-prune. This rule was run at all sparsity levels from $10\%$ to $80\%$.

\subsubsection*{Sample-and-rescale with uninformative probabilities}
To test whether sample-and-rescale preserves performance regardless of the retention probabilities, we applied it with probabilities that carry no information about individual edges. In the uniform-probability control, every edge is retained with probability equal to the target density and surviving weights are rescaled by the inverse of the target density. In the shuffled-probability control, the noise-prune probabilities $p_{ij}$ are randomly permuted across edges before sampling and rescaling, so that the distribution of probabilities and of rescale factors is preserved but their assignment to edges is destroyed. Both controls were run at all sparsity levels from $10\%$ to $80\%$.

\subsubsection*{Gain-matched baselines}
Because sample-and-rescale restores the expected recurrent gain lost to pruning while magnitude and random pruning do not, we also compared the methods at matched expected recurrent gain. In the density-matched baseline, magnitude or random pruning is applied and all surviving weights are multiplied by the inverse of the retained density. In the weight-matched baseline, surviving weights after magnitude pruning are multiplied by the constant that restores the total absolute recurrent weight $\sum_{ij}|w_{ij}|$ to its unpruned value (for random pruning this constant equals the inverse of the retained density in expectation, so the two baselines coincide). In two further baselines, surviving weights after magnitude pruning are multiplied, for each neuron, by the constant that restores the total absolute weight of that neuron's incoming connections, or, separately, of its outgoing connections. In all baselines the set of retained connections is unchanged and only the surviving weight magnitudes change. All baselines were run at all sparsity levels from $10\%$ to $80\%$.

\subsubsection*{Global rescaling with noise-prune scores}
To test whether the specific per-edge rescaling of noise-prune is key, rather than rescaling of any kind, we applied the same two global factors to networks pruned by sampling with the noise-prune probabilities. Both controls were run at all sparsity levels from $10\%$ to $80\%$.

\subsubsection*{Capped rescaling}
The rescaling in noise-prune amplifies surviving edges by $1/p_{ij}$. When $p_{ij}$ is small, this factor grows without bound, injecting high-variance perturbations into the sparse network. We introduce a capped rescale that bounds the amplification at a constant $c$, so that surviving edges are strengthened by $\min(1/p_{ij}, c)$ rather than $1/p_{ij}$. Capping under-compensates affected edges, trading exact operator preservation in expectation for bounded per-edge perturbation.

To choose $c$, we first swept a percentile cap that bounds the amplification at the $q$-th percentile of the amplification values $\{1/p_{ij}\}$ computed over all positive-probability candidate edges (i.e., not only the realized survivors), from $q = 10$ to $q = 90$ at each sparsity level from $50\%$ to $80\%$, with three pruning seeds per network and sparsity level. Lower $q$ caps more aggressively, and the uncapped rule corresponds to $q = 100$. Task retention peaked at intermediate percentiles at every sparsity level. We then set $c$ to the value of the best-performing percentile, averaged over networks, pruning seeds, and the $50\%$ to $80\%$ sparsity levels, and used this single constant at all sparsity levels. We carried out this procedure separately for noise-prune and sampled magnitude. For noise-prune, retention peaked at the 50th percentile (Fig.~\ref{fig:capped_task}A), giving $c = 3.695$. For sample-and-rescale with magnitude scores, retention peaked at the 60th percentile (Supplementary Fig.~S4), giving $c = 4.549$.

\subsubsection*{Optimal brain surgeon for recurrent activity preservation}
Consider pruning a network with parameters $\vc{w_0}$ while trying to preserve some loss function $L(\vc{w})$. If the network is near a minimum of the function $L$, then the gradient of $L$ at $\vc{w_0}$ is close to $0$ and the effect on $L$ of small perturbations of $\vc{w}$ around $\vc{w_0}$ is captured by the second-derivative or Hessian of $L$ at $\vc{w_0}$. The Optimal Brain Damage algorithm uses the Hessian to approximate the effect of removing each weight on the loss function, and then prunes the weights that have the least impact \cite{lecun90}. Instead of only considering perturbations that remove a weight, the Optimal Brain Surgeon algorithm instead uses the Hessian to approximate the effect of removing a weight while simultaneously adjusting the remaining weights to compensate \cite{hassibi93}. It then prunes the weights that have lowest impact while carrying out the predicted compensations. Note that carrying out this computation requires inverting the Hessian matrix, which is typically infeasible to do exactly. 

The layer-wise optimal brain surgeon algorithm applies OBS to preserve activity across each layer of a feedforward neural network, with the loss function chosen to be the mean-squared difference of pruned network activity from that of the unpruned neural network \cite{dong17}. In this setting, the Hessian has simple block-diagonal structure and can be expressed in terms of the unnormalized covariance (second moment) matrix of the inputs to the layer being pruned. 

We instead apply this layer-wise framework to the recurrent network (and call this recurrent-OBS in the Results), treating the recurrent map $r \to W_{\mathrm{rec}} r$ as a single layer whose inputs are the firing-rate states. Following the layer-wise OBS loss formulation\cite{dong17}, we approximate the recurrent layer Hessian using the damped second-moment matrix of these recurrent inputs. We define $\hat{S} = \frac{1}{M} X^\top X + \lambda I$, where the rows of $X \in \mathbb{R}^{M \times H}$ are the firing-rate states at $M = 25{,}000$ fixed calibration samples drawn from task input, and damping $\lambda = 10^{-3}$ stabilizes the inverse-Hessian estimate used for compensation. The Hessian is then approximately block-diagonal with $N$ copies of $\hat{S}$ along the diagonal (i.e., $\mathcal{H} = I_N \otimes \hat{S}$, where $I_N$ is the $N \times N$ identity matrix and $\otimes$ is the Kronecker product) and the inverse-Hessian is similarly block-diagonal with the inverse of $\hat{S}$.

Each edge is ranked by its OBS sensitivity (squared weight divided by the corresponding inverse-Hessian diagonal entry), and the lowest-saliency edges are removed in a single pass to reach the target density. Surviving weights then receive the closed-form OBS compensation update, which redistributes the pruned drive onto correlated surviving inputs through the off-diagonal inverse Hessian. We also apply this compensation in a single pass and do not recalculate the Hessian after each weight is compensated for. 

Recurrent-OBS is thus non-local and more computationally-intensive than noise-prune. It is unsupervised and does not require access to the network's task performance or loss function. Instead, like noise-prune it extracts the importance of connections from the population activity covariance matrix. However, unlike noise-prune, it requires inverting the covariance matrix and the compensation step requires changing weights other than the weight being pruned.

\subsection*{Evaluation pipeline}
We evaluate all pruning methods on all 24 networks at eight pruning levels ($10\%$ to $80\%$ in steps of $10\%$). Stochastic methods are run with three separate pruning seeds and deterministic methods are run once. Each pruned network is evaluated on task preservation. Results are aggregated across pruning seeds, model seeds, and tasks as described under Statistical comparisons, and figure error bars show 68\% bootstrap confidence intervals.

For each pruned network, we evaluate on a set of 128 batches, held constant across all methods, pruning levels, and seeds for that task. The measured metric is post-fixation sequence accuracy (\texttt{post\_acc\_sequence}): the fraction of valid (timestep, batch element) pairs where the predicted class matches the target. We report this as task retention: the pruned network's post-fixation sequence accuracy divided by that of the unpruned baseline. Chance accuracy is $1/15 \approx 0.067$ on every task, since each scored position is classified into one of $15$ target classes (Supplementary Table~S1). In units of task retention, chance corresponds to $0.075$ to $0.114$ depending on the task's unpruned accuracy (Supplementary Table~S2), and is marked in Fig.~\ref{fig:task}. As a supplementary check, we also report floor-corrected retention, $(\mathrm{acc}_{\mathrm{pruned}} - \mathrm{chance})/(\mathrm{acc}_{\mathrm{unpruned}} - \mathrm{chance})$, which is $0$ at chance and $1$ for the unpruned network (Supplementary Table~S3 and Supplementary Fig.~S3).

Note that these networks seek to output a sequence of results, with the timing of the sequence being more sensitive to pruning than the elements of the sequence. Thus, pruning rules can show below chance performance post-pruning if they output the correct sequence but shifted in time. This effect is slightly visible in the performance of the random pruning in Fig.~\ref{fig:task}.

Alongside task retention, we report the spectral abscissa of each pruned recurrent matrix as a diagnostic of dynamical stability, defined as $a(W_{\mathrm{rec}}) = \max_i \mathrm{Re}(\lambda_i)$, the largest real part among the eigenvalues of $W_{\mathrm{rec}}$. Fig.~\ref{fig:effect_of_uncapped_rescale}B reports the change in spectral abscissa from the unpruned network, $a(W^{sparse}_{\mathrm{rec}}) - a(W_{\mathrm{rec}})$. Because the linearized dynamics are governed by $A = W_{\mathrm{rec}} - I$, whose eigenvalues are those of $W_{\mathrm{rec}}$ shifted down by one, $a(W_{\mathrm{rec}}) < 1$ means every eigenvalue of $A$ has negative real part and perturbations about the origin decay, so the resting state is stable. When $a(W_{\mathrm{rec}})$ exceeds 1, at least one mode of $A$ has a positive growth rate and the origin is no longer attracting, so activity grows rather than settling. The tanh nonlinearity nonetheless bounds the state, so this growth does not diverge but instead moves the network onto other bounded dynamics.

\subsection*{Statistical comparisons}
Comparisons between methods are made across the 8 tasks ($n = 8$). Pruning seeds and then network seeds are averaged within each task, giving one retention curve per task and method (shown for every task in Supplementary Fig.~S2). The mean of these over tasks is the retention curve shown in the main text figures. To summarize performance across the pruning range, we compute the area under each task's retention curve from $10\%$ to $80\%$ sparsity by the trapezoid rule, normalized so that perfect retention across the range gives a score of $1$. We average these measures over tasks when reporting aggregate performance. The effect size for a comparison between two methods is the mean over tasks of the task-by-task difference in this area, which is directly interpretable as the average improvement in retention across the pruning range. We also report the task-by-task and aggregate difference in retention at each sparsity, and give both types of comparison for noise-prune against the other methods in Supplementary Tables~S5 and~S8.

Confidence intervals were obtained by hierarchical bootstrap, resampling the 8 tasks with replacement and then resampling the 3 network seeds with replacement within each sampled task, with the same resampled networks used for every method so that comparisons remain paired. All measures were recomputed on each of 10,000 surrogate datasets. Reported 95\% confidence intervals are the 2.5th and 97.5th percentiles of the resulting distributions, and figure error bars show 68\% intervals, the 16th and 84th percentiles. Significance was assessed with a two-sided exact sign-flip test on the eight task-by-task differences, enumerating all $2^8 = 256$ sign assignments. Differences are called significant at $p<0.05$. The smallest attainable two-sided $p$-value is $2/256 \approx 0.0078$, reached when all eight differences share a sign, so we report the number of tasks favoring each method alongside every $p$-value. For almost all comparisons of interest the differences are positive for every task and the $p$-value is at this floor. For the comparisons of noise-prune against every other method (Supplementary Table~S5) and of each swept cap against the uncapped rule (Supplementary Table~S7), we also report family-wise corrected $p$-values from the Westfall--Young step-down min-$P$ procedure\cite{westfall1993}, computed over the same $256$ sign assignments.

\subsection*{Implementation and reproducibility}
All experiments use Python 3.13 with PyTorch 2.8\cite{paszke2019pytorch}, NumPy~\texttt{2.1.3}, SciPy~\texttt{1.14.1}, and the Mod-Cog task generator~\cite{khona2023modcog} (\url{https://github.com/mikailkhona/Mod_Cog}). Each task is trained with three independently seeded networks (different weight initializations and training-data orders). Within each of the three pruning seeds, random pruning draws from the PyTorch global generator (\texttt{config.seed}), while noise injection (noise-prune) and Bernoulli sampling (rescale variants) draw from a separate NumPy generator (\texttt{noise\_rng\_seed}); the pruning seed also sets the calibration batches used for scoring (\texttt{score\_batch\_seed}). Scoring and evaluation use fixed, persisted batches drawn from seeds disjoint from the training-seed space (score-batch seeds 100{,}000--100{,}002, evaluation-batch seed 200{,}000), and validation during training used deterministic batches (eval seed 0). 

All figures were produced by the authors using R version 4.5.1 with ggplot2, \url{https://ggplot2.tidyverse.org}.

\subsection*{Code Availability}
The custom code implementing the pruning methods and all analyses is available in the Zenodo repository: \url{https://doi.org/10.5281/zenodo.23653030}.

\subsection*{AI-Use Declaration}
The authors used Large Language Models (GPT $5.4$, GPT $6$, and Opus $4.7$) to assist with implementing and debugging code and editing text. All AI-assisted output was reviewed, verified, and edited by the authors, who take full responsibility for the content and accuracy of the code and conclusions. The authors retained full control over the study design, analyses, interpretation of results, and manuscript content.

\section*{Data availability}
The datasets generated and analyzed during the study are available in the Zenodo repository, \url{https://doi.org/10.5281/zenodo.23653028}.

\section*{Funding}
This study is based upon work supported by the Air Force Office of Scientific Research (AFOSR) under award number FA9550-22-1-0532. 

\section*{Acknowledgements}
The authors are grateful to R. Guy, T. Lewis, S. Schreiber and the California State Summer School for Mathematics \& Science (COSMOS) for helping initiate this collaboration.

\section*{Author contributions statement}
S.S. and R.C. designed the study. S.S. implemented and trained the models, performed the numerical experiments, and analyzed results. S.S. and R.C. interpreted results and wrote the manuscript. 

\section*{Additional information}
\textbf{Competing interests} The authors declare no competing interests.

\end{document}


\begin{center}
{\LARGE\bfseries Supplementary Information}\\[8pt]
\end{center}
\suppressfloats[t]

\begin{figure}[!htb]\centering
\includegraphics[width=\textwidth]{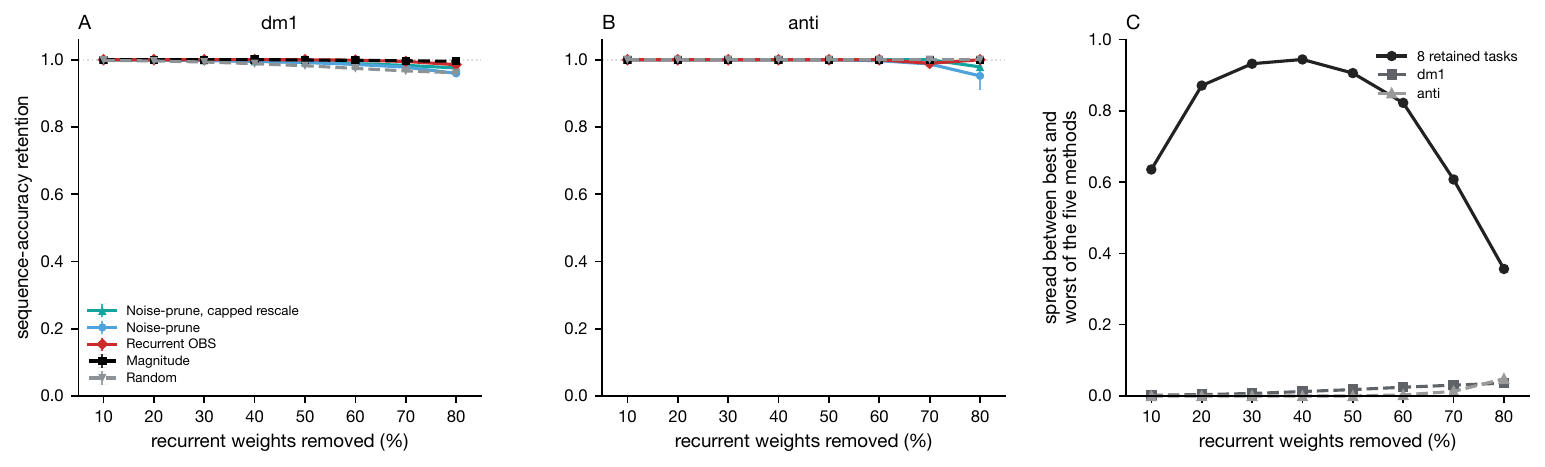}
\caption{\textbf{Task retention on the excluded dm1 and anti tasks.} Task retention of networks trained on dm1 (\textbf{A}) and anti (\textbf{B}) under the same protocol as the eight retained tasks ($n = 3$ networks each), plotted against the percentage of recurrent weights pruned, for the four methods of Fig.~2 and capped noise-prune (Fig.~5). (\textbf{C}) Difference in retention between the best and worst of the five methods at each sparsity, for the eight retained tasks and for dm1 and anti. Points show the mean and error bars the $68\%$ bootstrap interval, over tasks for the retained tasks and over networks for dm1 and anti.}
\label{fig:S_excluded}
\end{figure}

\begin{figure}[!htb]\centering
\includegraphics[width=\textwidth]{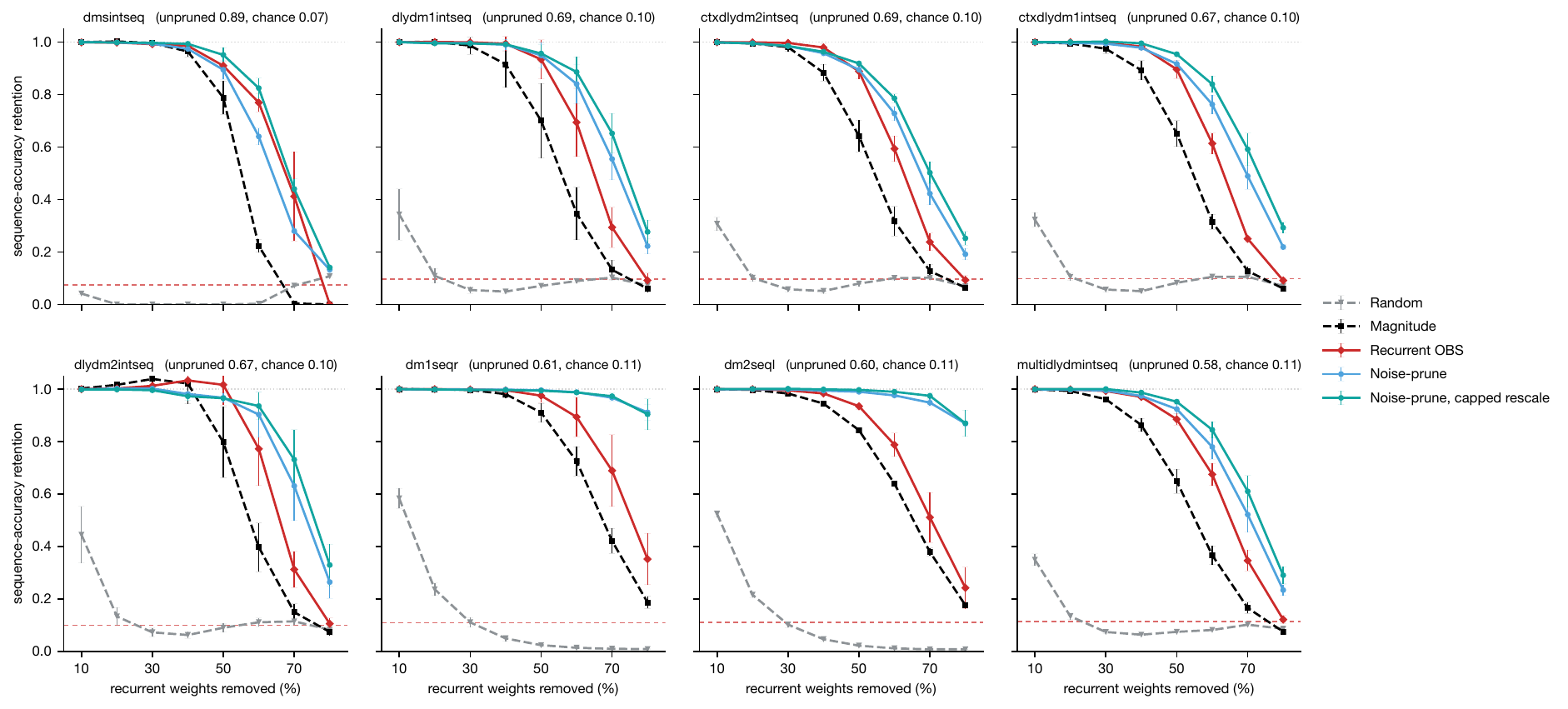}
\caption{\textbf{Task retention on each of the eight tasks.} Task retention for each task separately, plotted against the percentage of recurrent weights pruned, for the four methods of Fig.~2 and capped noise-prune (Fig.~5). Each panel gives the task name, its mean unpruned accuracy over the three network seeds, and the chance level in units of retention, which is marked by the dashed line. Points show the mean over the three networks trained on that task, with stochastic pruning methods averaged across three pruning seeds within each network, and error bars the $68\%$ bootstrap interval across networks.}
\label{fig:S_pertask}
\end{figure}

\begin{figure}[!htb]\centering
\includegraphics[width=0.62\textwidth]{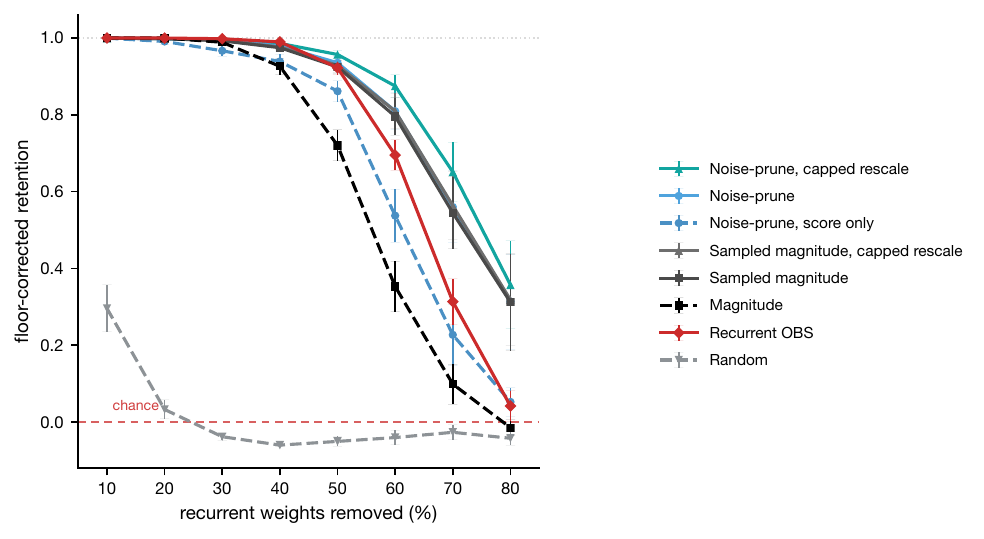}
\caption{\textbf{Floor-corrected task retention across pruning methods and sparsity levels.} Floor-corrected retention, $(\mathrm{acc}_{\mathrm{pruned}} - \mathrm{chance})/(\mathrm{acc}_{\mathrm{unpruned}} - \mathrm{chance})$, which is $1$ for the unpruned network and $0$ at chance (dashed line), plotted against the percentage of recurrent weights pruned for the methods of Figs.~2, 3A, 3D and~5B. Points show the mean over tasks and error bars the $68\%$ bootstrap interval. Values are given in Supplementary Table~\ref{tab:S_floor}.}
\label{fig:S_floor}
\end{figure}

\begin{figure}[!htb]\centering
\includegraphics[width=0.6\textwidth]{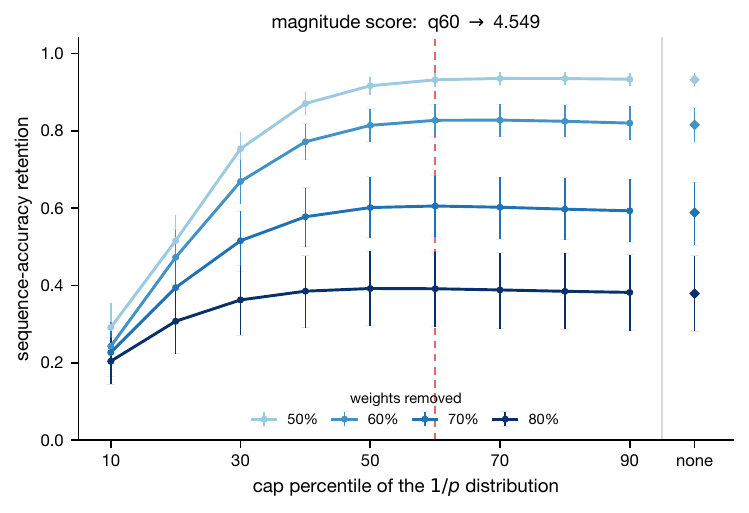}
\caption{\textbf{Effect of capping rescale factors on task retention for sampled magnitude.} Task retention when per-edge amplification is capped at the $q$th percentile of the candidate-edge amplification distribution, for $q = 10$ to $90$ at each sparsity level from $50\%$ to $80\%$, with the uncapped rule shown at right. The dashed line marks the percentile at which retention pooled over the four sparsity levels peaks ($q = 60$), from which the constant cap $c = 4.549$ is derived as described in Methods. Points show the mean over tasks and error bars the $68\%$ bootstrap interval. The corresponding sweep for noise-prune is shown in Fig.~5A.}
\label{fig:S_magsweep}
\end{figure}

\begin{figure}[!htb]\centering
\includegraphics[width=0.95\textwidth]{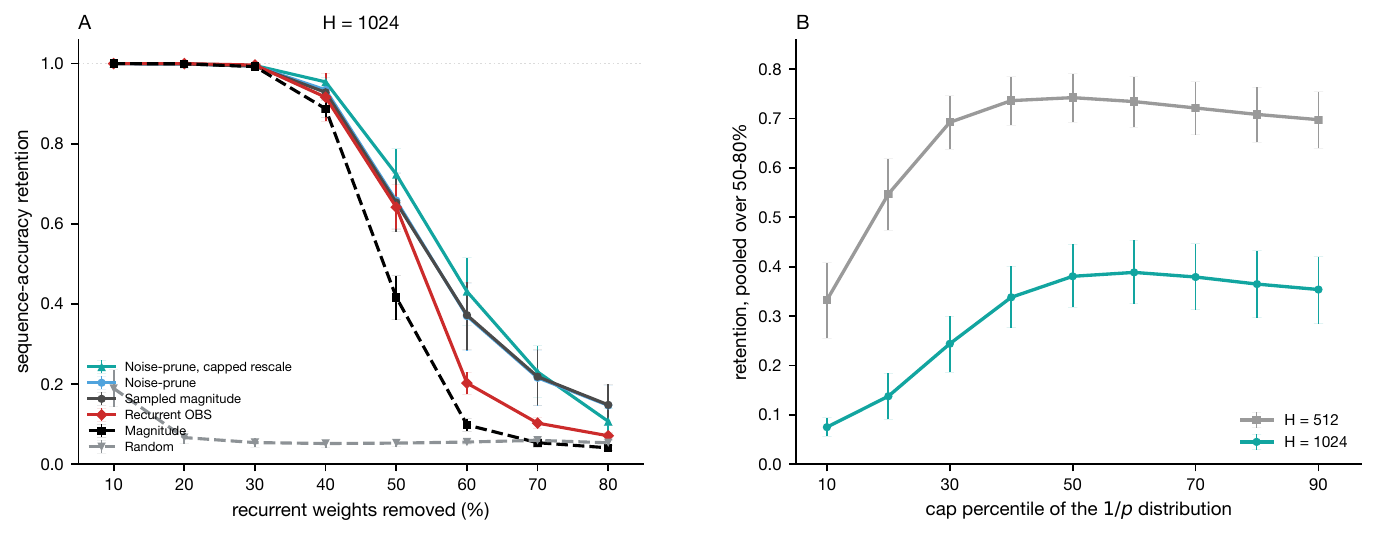}
\caption{\textbf{Task retention and cap-percentile sweep at $H = 1024$.} (\textbf{A}) Task retention of networks with hidden size $H = 1024$ (8 tasks $\times$ 3 network seeds, trained as in Methods), plotted against the percentage of recurrent weights removed, for capped noise-prune, noise-prune, sampled magnitude, recurrent OBS, magnitude pruning, and random pruning. Capped noise-prune uses the constant $c = 3.695$ derived at $H = 512$. (\textbf{B}) Task retention for noise-prune when per-edge amplification is capped at the $q$th percentile of the candidate-edge amplification distribution, for $q = 10$ to $90$, pooled over the $50\%$ to $80\%$ sparsity levels with three pruning seeds per network and sparsity level, at $H = 1024$ and, for comparison, at $H = 512$ (the sweep of Fig.~5A). Retention peaks at $q = 60$ at $H = 1024$ and at $q = 50$ at $H = 512$, but the two percentiles are not distinguishable at $H = 1024$ (difference in pooled retention of $0.008$, 95\% CI $[-0.011, 0.024]$, $6/8$ tasks, $p = 0.38$), so the optimal cap percentile does not change appreciably with network size. Points show the mean over tasks and error bars the $68\%$ bootstrap interval.}
\label{fig:S_h1024}
\end{figure}

\clearpage

\begin{table}[!htb]\centering
\caption{\textbf{Specifications of the eight Mod-Cog tasks used in this study\cite{khona2023modcog}.} Tasks were generated with the Mod-Cog builder. Trial periods list the epoch sequence and timing\,(ms) gives epoch durations (fixation durations sampled uniformly where indicated). All tasks use an environment timestep of $100$\,ms and a $16$-point ring representation. Outputs are a fixation unit and the $16$ ring units, with ring position $0$ serving as the fixation label, so responses are scored over the remaining $15$ ring positions.}
\label{tab:S1_tasks}\footnotesize
\begin{tabular}{@{}llp{3.4cm}p{3.9cm}@{}}\toprule
Task & Family & Trial periods & Timing (ms) \\\midrule
\texttt{ctxdlydm2intseq} & delayed decision making with context & fixation; stim1; delay; stim2; decision & fixation uniform 200-500; stim1 500; delay 500; stim2 500; decision 1000 \\
\texttt{ctxdlydm1intseq} & delayed decision making with context & fixation; stim1; delay; stim2; decision & fixation uniform 200-500; stim1 500; delay 500; stim2 500; decision 1000 \\
\texttt{dlydm1intseq} & delayed decision making & fixation; stim1; delay; stim2; decision & fixation uniform 200-500; stim1 500; delay 500; stim2 500; decision 1000 \\
\texttt{dlydm2intseq} & delayed decision making & fixation; stim1; delay; stim2; decision & fixation uniform 200-500; stim1 500; delay 500; stim2 500; decision 1000 \\
\texttt{multidlydmintseq} & multimodal delayed decision making & fixation; stim1; delay; stim2; decision & fixation uniform 200-500; stim1 500; delay 500; stim2 500; decision 1000 \\
\texttt{dm1seqr} & decision making & fixation; stimulus; decision & fixation uniform 200-500; stimulus 500; decision 1000 \\
\texttt{dm2seql} & decision making & fixation; stimulus; decision & fixation uniform 200-500; stimulus 500; decision 1000 \\
\texttt{dmsintseq} & delayed match to sample & fixation; sample; delay; test; decision & fixation 300; sample 500; delay 500; test 500; decision 1000 \\
\bottomrule\end{tabular}\end{table}

\begin{table}[!htb]\centering
\caption{\textbf{Unpruned performance and chance level for each task.} Post-fixation sequence accuracy of the unpruned networks on the $128$ held-out evaluation batches, as the mean, standard deviation, minimum and maximum over the three network seeds. Chance accuracy is $1/15$ on every task, and chance in units of task retention is chance accuracy divided by the mean unpruned accuracy.}
\label{tab:S_chance}\footnotesize
\begin{tabular}{lcccccc}\toprule
& \multicolumn{4}{c}{Unpruned sequence accuracy} & & \\
\cmidrule(lr){2-5}
Task & Mean & s.d. & Min & Max & Chance accuracy & Chance in retention units \\\midrule
\texttt{ctxdlydm1intseq} & 0.672 & 0.059 & 0.609 & 0.725 & 0.067 & 0.099 \\
\texttt{ctxdlydm2intseq} & 0.689 & 0.034 & 0.655 & 0.723 & 0.067 & 0.097 \\
\texttt{dlydm1intseq} & 0.693 & 0.077 & 0.609 & 0.760 & 0.067 & 0.096 \\
\texttt{dlydm2intseq} & 0.670 & 0.098 & 0.557 & 0.732 & 0.067 & 0.099 \\
\texttt{dm1seqr} & 0.608 & 0.025 & 0.583 & 0.633 & 0.067 & 0.110 \\
\texttt{dm2seql} & 0.597 & 0.039 & 0.553 & 0.624 & 0.067 & 0.112 \\
\texttt{dmsintseq} & 0.892 & 0.032 & 0.864 & 0.927 & 0.067 & 0.075 \\
\texttt{multidlydmintseq} & 0.582 & 0.023 & 0.569 & 0.608 & 0.067 & 0.114 \\
\bottomrule\end{tabular}\end{table}

\begin{table}[!htb]\centering
\caption{\textbf{Floor-corrected retention by method and sparsity.} Mean over tasks of floor-corrected retention, $(\mathrm{acc}_{\mathrm{pruned}} - \mathrm{chance})/(\mathrm{acc}_{\mathrm{unpruned}} - \mathrm{chance})$, with $95\%$ bootstrap confidence intervals in brackets and the number of the $24$ networks at or below chance in parentheses. Plotted in Supplementary Fig.~\ref{fig:S_floor}.}
\label{tab:S_floor}\tiny
\setlength{\tabcolsep}{1.5pt}
\begin{tabular}{lcccc}\toprule
Method & 10\% & 20\% & 30\% & 40\% \\\midrule
Noise-prune, capped rescale & 1.000 [0.999, 1.000] (0) & 0.998 [0.996, 1.000] (0) & 0.997 [0.992, 1.001] (0) & 0.986 [0.975, 0.996] (0) \\
Noise-prune, rescale & 1.000 [0.999, 1.000] (0) & 0.998 [0.996, 1.000] (0) & 0.995 [0.990, 0.999] (0) & 0.979 [0.968, 0.990] (0) \\
Noise-prune, score only & 1.000 [0.995, 1.004] (0) & 0.991 [0.977, 1.005] (0) & 0.967 [0.936, 0.994] (0) & 0.939 [0.896, 0.971] (0) \\
Sampled magnitude, capped rescale & 0.999 [0.999, 1.000] (0) & 0.999 [0.997, 1.001] (0) & 0.994 [0.988, 0.998] (0) & 0.977 [0.964, 0.989] (0) \\
Sampled magnitude & 1.000 [0.999, 1.000] (0) & 0.998 [0.997, 1.000] (0) & 0.993 [0.987, 0.997] (0) & 0.975 [0.962, 0.987] (0) \\
Magnitude & 1.000 [0.999, 1.001] (0) & 1.000 [0.995, 1.007] (0) & 0.989 [0.971, 1.011] (0) & 0.926 [0.880, 0.978] (0) \\
Recurrent OBS & 1.000 [1.000, 1.000] (0) & 1.000 [0.999, 1.001] (0) & 0.998 [0.993, 1.004] (0) & 0.990 [0.975, 1.009] (0) \\
Random & 0.296 [0.176, 0.407] (3) & 0.033 [-0.011, 0.079] (8) & -0.038 [-0.054, -0.021] (22) & -0.060 [-0.069, -0.051] (24) \\
\midrule
Method & 50\% & 60\% & 70\% & 80\% \\\midrule
Noise-prune, capped rescale & 0.957 [0.936, 0.978] (0) & 0.875 [0.817, 0.935] (0) & 0.651 [0.515, 0.802] (0) & 0.357 [0.170, 0.585] (0) \\
Noise-prune, rescale & 0.936 [0.903, 0.968] (0) & 0.810 [0.722, 0.902] (0) & 0.560 [0.396, 0.743] (0) & 0.314 [0.116, 0.577] (0) \\
Noise-prune, score only & 0.862 [0.802, 0.915] (0) & 0.538 [0.406, 0.677] (0) & 0.227 [0.090, 0.388] (3) & 0.052 [-0.011, 0.130] (7) \\
Sampled magnitude, capped rescale & 0.929 [0.893, 0.965] (0) & 0.810 [0.726, 0.900] (0) & 0.564 [0.410, 0.742] (0) & 0.318 [0.127, 0.569] (0) \\
Sampled magnitude & 0.925 [0.888, 0.962] (0) & 0.796 [0.707, 0.892] (0) & 0.545 [0.385, 0.731] (0) & 0.312 [0.110, 0.584] (0) \\
Magnitude & 0.721 [0.632, 0.810] (0) & 0.353 [0.239, 0.486] (0) & 0.099 [0.012, 0.200] (4) & -0.015 [-0.050, 0.027] (18) \\
Recurrent OBS & 0.923 [0.884, 0.971] (0) & 0.695 [0.612, 0.785] (0) & 0.314 [0.203, 0.448] (0) & 0.042 [-0.023, 0.129] (11) \\
Random & -0.050 [-0.075, -0.027] (23) & -0.041 [-0.075, -0.008] (17) & -0.027 [-0.066, 0.005] (14) & -0.042 [-0.074, -0.012] (21) \\
\bottomrule\end{tabular}\end{table}

\begin{table}[!htb]\centering
\caption{\textbf{Task retention of noise-prune under alternative covariance estimates.} Retention of noise-prune (rescale) with the covariance estimated as in Methods (final-timestep states under task input with injected noise at $\sigma = \sigma_{\mathrm{nat}}$) and with four variants: the injected noise removed ($\sigma = 0.001\,\sigma_{\mathrm{nat}}$), the task input removed, states sampled from all timesteps of each rollout rather than the final timestep, and all timesteps with the task input removed. Difference is the mean over tasks of the task-by-task difference (Methods estimate minus variant), with its $95\%$ confidence interval, the number of tasks favoring the Methods estimate, and the exact sign-flip $p$-value. The all-timestep variants were run with one pruning seed per network.}
\label{tab:S_covariance}\footnotesize
\setlength{\tabcolsep}{4pt}
\begin{tabular}{llcccccc}\toprule
Variant & Sparsity & Main text & Variant & Difference & 95\% CI & Tasks & $p$ \\\midrule
No injected noise & 10\% & 1.000 & 0.998 & +0.002 & [-0.000, +0.004] & 6/8 & 0.156 \\
 & 20\% & 0.998 & 0.983 & +0.015 & [+0.002, +0.039] & 6/8 & 0.0547 \\
 & 30\% & 0.996 & 0.944 & +0.052 & [+0.008, +0.135] & 8/8 & 0.0078 \\
 & 40\% & 0.981 & 0.881 & +0.100 & [+0.014, +0.240] & 8/8 & 0.0078 \\
 & 50\% & 0.942 & 0.778 & +0.164 & [+0.048, +0.337] & 8/8 & 0.0078 \\
 & 60\% & 0.827 & 0.619 & +0.208 & [+0.103, +0.330] & 8/8 & 0.0078 \\
 & 70\% & 0.602 & 0.436 & +0.166 & [+0.099, +0.225] & 8/8 & 0.0078 \\
 & 80\% & 0.381 & 0.298 & +0.083 & [+0.063, +0.103] & 8/8 & 0.0078 \\
 & AUC 10--80\% & 0.862 & 0.756 & +0.107 & [+0.048, +0.188] & 8/8 & 0.0078 \\
\midrule
No task input & 10\% & 1.000 & 1.000 & +0.000 & [-0.000, +0.000] & 3/8 & 0.953 \\
 & 20\% & 0.998 & 0.999 & -0.000 & [-0.001, +0.000] & 2/8 & 0.164 \\
 & 30\% & 0.996 & 0.995 & +0.001 & [-0.001, +0.003] & 5/8 & 0.281 \\
 & 40\% & 0.981 & 0.982 & -0.001 & [-0.005, +0.003] & 4/8 & 0.719 \\
 & 50\% & 0.942 & 0.940 & +0.002 & [-0.005, +0.008] & 4/8 & 0.500 \\
 & 60\% & 0.827 & 0.824 & +0.003 & [-0.010, +0.013] & 6/8 & 0.531 \\
 & 70\% & 0.602 & 0.600 & +0.002 & [-0.015, +0.018] & 5/8 & 0.852 \\
 & 80\% & 0.381 & 0.378 & +0.002 & [-0.002, +0.007] & 5/8 & 0.281 \\
 & AUC 10--80\% & 0.862 & 0.861 & +0.001 & [-0.004, +0.005] & 5/8 & 0.562 \\
\midrule
All timesteps & 10\% & 1.000 & 1.000 & -0.000 & [-0.001, +0.000] & 2/8 & 0.312 \\
 & 20\% & 0.998 & 0.999 & -0.000 & [-0.002, +0.002] & 4/8 & 0.805 \\
 & 30\% & 0.996 & 0.994 & +0.002 & [-0.002, +0.005] & 5/8 & 0.328 \\
 & 40\% & 0.981 & 0.979 & +0.003 & [-0.007, +0.013] & 6/8 & 0.617 \\
 & 50\% & 0.942 & 0.935 & +0.007 & [-0.011, +0.022] & 5/8 & 0.156 \\
 & 60\% & 0.827 & 0.817 & +0.010 & [-0.007, +0.030] & 5/8 & 0.211 \\
 & 70\% & 0.602 & 0.587 & +0.015 & [-0.018, +0.049] & 5/8 & 0.320 \\
 & 80\% & 0.381 & 0.385 & -0.004 & [-0.023, +0.011] & 3/8 & 0.602 \\
 & AUC 10--80\% & 0.862 & 0.857 & +0.005 & [-0.005, +0.015] & 6/8 & 0.289 \\
\midrule
All timesteps, no task input & 10\% & 1.000 & 1.000 & -0.000 & [-0.001, +0.001] & 3/8 & 0.695 \\
 & 20\% & 0.998 & 0.998 & -0.000 & [-0.002, +0.002] & 4/8 & 0.945 \\
 & 30\% & 0.996 & 0.993 & +0.003 & [-0.002, +0.009] & 5/8 & 0.398 \\
 & 40\% & 0.981 & 0.979 & +0.002 & [-0.008, +0.012] & 4/8 & 0.734 \\
 & 50\% & 0.942 & 0.941 & +0.001 & [-0.016, +0.014] & 4/8 & 0.883 \\
 & 60\% & 0.827 & 0.838 & -0.011 & [-0.034, +0.012] & 3/8 & 0.336 \\
 & 70\% & 0.602 & 0.608 & -0.006 & [-0.037, +0.022] & 3/8 & 0.680 \\
 & 80\% & 0.381 & 0.396 & -0.015 & [-0.032, -0.001] & 1/8 & 0.0234 \\
 & AUC 10--80\% & 0.862 & 0.865 & -0.003 & [-0.014, +0.008] & 4/8 & 0.578 \\
\bottomrule\end{tabular}\end{table}

\begin{table}[!htb]\centering
\caption{\textbf{Comparisons of noise-prune to the other methods across the pruning range.} Effect size is the mean over tasks of the task-by-task difference in normalized area under the retention curve (first method minus second), with its $95\%$ bootstrap confidence interval, the number of tasks favoring the first method, the exact two-sided sign-flip $p$-value ($n = 8$ tasks, smallest attainable value $2/256 \approx 0.0078$), and the family-wise corrected $p$-value from the Westfall--Young step-down procedure over the same sign assignments, within each family (seven comparisons for capped noise-prune, six for the original rule).}
\label{tab:S_main}\scriptsize
\setlength{\tabcolsep}{4pt}
\begin{tabular}{llccccc}\toprule
Method & Compared to & Effect & 95\% CI & Tasks & $p$ & Corrected $p$ \\\midrule
Noise-prune, capped rescale & Noise-prune, rescale & +0.027 & [+0.014, +0.040] & 7/8 & 0.0156 & 0.0156 \\
 & Noise-prune, score only & +0.141 & [+0.122, +0.163] & 8/8 & 0.0078 & 0.0156 \\
 & Sampled magnitude, capped rescale & +0.027 & [+0.015, +0.039] & 8/8 & 0.0078 & 0.0156 \\
 & Sampled magnitude & +0.033 & [+0.018, +0.047] & 7/8 & 0.0156 & 0.0156 \\
 & Magnitude & +0.201 & [+0.183, +0.218] & 8/8 & 0.0078 & 0.0156 \\
 & Recurrent OBS & +0.090 & [+0.065, +0.115] & 8/8 & 0.0078 & 0.0156 \\
 & Random & +0.796 & [+0.759, +0.837] & 8/8 & 0.0078 & 0.0156 \\
\midrule
Noise-prune, rescale & Noise-prune, score only & +0.114 & [+0.095, +0.133] & 8/8 & 0.0078 & 0.0312 \\
 & Sampled magnitude, capped rescale & +0.000 & [-0.007, +0.007] & 5/8 & 0.891 & 0.891 \\
 & Sampled magnitude & +0.006 & [-0.001, +0.013] & 6/8 & 0.117 & 0.195 \\
 & Magnitude & +0.174 & [+0.154, +0.193] & 8/8 & 0.0078 & 0.0312 \\
 & Recurrent OBS & +0.063 & [+0.029, +0.096] & 7/8 & 0.0156 & 0.0469 \\
 & Random & +0.769 & [+0.726, +0.818] & 8/8 & 0.0078 & 0.0312 \\
\bottomrule\end{tabular}\end{table}

\begin{table}[!htb]\centering
\caption{\textbf{Comparisons between the components of noise-prune and its controls.} For each comparison, the average task retention across the pruning range ($10\%$ to $80\%$) of the first and second method is given with its $95\%$ bootstrap confidence interval, followed by the effect size, the mean over tasks of the task-by-task difference in average retention (first method minus second), with its $95\%$ confidence interval, the number of tasks favoring the first method, and the exact sign-flip $p$-value. The second block gives the uninformative-probability controls of Fig.~3. The third block gives the global-rescaling controls of Fig.~3, applied to the connections sampled by noise-prune in place of its per-edge rescaling and to magnitude and random pruning, in which surviving weights are multiplied by the inverse of the retained density (density-matched), by the constant that restores the total absolute recurrent weight (weight-matched), or, for magnitude pruning, by the per-neuron constants that restore each neuron's total incoming or outgoing absolute weight (incoming and outgoing weight-matched); for random pruning the two coincide in expectation, and the sampled rule is sample-and-rescale with uniform probabilities. The noise-prune global-rescaling controls were run with one pruning seed per network. The fourth block compares each capped rule of Fig.~5 under its constant cap with the same rule under a per-network cap at the percentile from which the constant was derived, with retention pooled over the $50\%$ to $80\%$ sparsity levels.}
\label{tab:S_mechanism}\scriptsize
\setlength{\tabcolsep}{3pt}
\begin{tabular}{>{\raggedright\arraybackslash}p{4.4cm}cccccc}\toprule
Comparison & First method & Second method & Effect & 95\% CI & Tasks & $p$ \\\midrule
Noise-prune score only vs magnitude & 0.748 [0.698, 0.803] & 0.689 [0.647, 0.736] & +0.060 & [+0.032, +0.086] & 8/8 & 0.0078 \\
Sampled magnitude vs magnitude & 0.856 [0.806, 0.914] & 0.689 [0.647, 0.736] & +0.168 & [+0.150, +0.186] & 8/8 & 0.0078 \\
Noise-prune, rescale vs sampled magnitude & 0.862 [0.812, 0.918] & 0.856 [0.806, 0.914] & +0.006 & [-0.001, +0.013] & 6/8 & 0.117 \\
Noise-prune, capped rescale vs noise-prune, rescale & 0.889 [0.849, 0.934] & 0.862 [0.812, 0.918] & +0.027 & [+0.014, +0.040] & 7/8 & 0.0156 \\
Sampled magnitude, capped rescale vs sampled magnitude & 0.862 [0.814, 0.916] & 0.856 [0.806, 0.914] & +0.006 & [+0.002, +0.009] & 6/8 & 0.0312 \\
\midrule
Noise-prune, rescale vs shuffled probabilities & 0.862 [0.812, 0.918] & 0.342 [0.196, 0.531] & +0.520 & [+0.386, +0.623] & 8/8 & 0.0078 \\
Noise-prune, rescale vs uniform probabilities & 0.862 [0.812, 0.918] & 0.652 [0.540, 0.791] & +0.211 & [+0.127, +0.282] & 8/8 & 0.0078 \\
\midrule
Noise-prune, rescale vs noise-prune, weight-matched & 0.862 [0.812, 0.918] & 0.302 [0.254, 0.345] & +0.561 & [+0.520, +0.601] & 8/8 & 0.0078 \\
Noise-prune, rescale vs noise-prune, density-matched & 0.862 [0.812, 0.918] & 0.092 [0.072, 0.109] & +0.770 & [+0.719, +0.830] & 8/8 & 0.0078 \\
Sampled magnitude vs magnitude, weight-matched & 0.856 [0.806, 0.914] & 0.352 [0.300, 0.395] & +0.504 & [+0.456, +0.553] & 8/8 & 0.0078 \\
Sampled magnitude vs magnitude, density-matched & 0.856 [0.806, 0.914] & 0.081 [0.056, 0.103] & +0.775 & [+0.723, +0.833] & 8/8 & 0.0078 \\
Sampled magnitude vs magnitude, incoming weight-matched & 0.856 [0.806, 0.914] & 0.351 [0.293, 0.401] & +0.506 & [+0.461, +0.551] & 8/8 & 0.0078 \\
Sampled magnitude vs magnitude, outgoing weight-matched & 0.856 [0.806, 0.914] & 0.358 [0.299, 0.412] & +0.498 & [+0.453, +0.543] & 8/8 & 0.0078 \\
Random, uniform probabilities vs random, density-matched & 0.652 [0.540, 0.791] & 0.645 [0.532, 0.787] & +0.006 & [-0.009, +0.025] & 5/8 & 0.578 \\
\midrule
Noise-prune, constant cap $c=3.695$ vs percentile cap $q=50$ (50--80\%) & 0.738 [0.649, 0.841] & 0.742 [0.653, 0.846] & -0.004 & [-0.013, +0.002] & 3/8 & 0.352 \\
Sampled magnitude, constant cap $c=4.549$ vs percentile cap $q=60$ (50--80\%) & 0.688 [0.585, 0.810] & 0.689 [0.583, 0.813] & -0.001 & [-0.004, +0.003] & 4/8 & 0.727 \\
\bottomrule\end{tabular}\end{table}

\begin{table}[!htb]\centering
\caption{\textbf{Capped versus uncapped rescale comparisons.} Task retention when amplification is capped at the $q$th percentile of the candidate-edge amplification distribution, at each sparsity from $50\%$ to $80\%$ and pooled over the four levels, for noise-prune (Fig.~5A) and for sampled magnitude (Supplementary Fig.~\ref{fig:S_magsweep}). The constant $c$ is the cap value at that percentile averaged over networks, pruning seeds and the four sparsity levels. The last five columns compare each cap to the uncapped rule by the task-by-task difference in pooled retention, with its $95\%$ confidence interval, the number of tasks favoring the cap, the exact sign-flip $p$-value, and the family-wise corrected $p$-value from the Westfall--Young step-down procedure over the same sign assignments, treating the nine caps swept for each rule as one family.}
\label{tab:S_capsweep}\scriptsize
\setlength{\tabcolsep}{2.2pt}
\begin{tabular}{lcccccccccccc}\toprule
Cap & $c$ & 50\% & 60\% & 70\% & 80\% & Pooled & Cap minus uncapped & 95\% CI & Tasks & $p$ & Corrected $p$ \\\midrule
\multicolumn{12}{l}{\textit{Noise-prune}} \\
$q=10$ & 1.590 & 0.433 & 0.321 & 0.304 & 0.272 & 0.332 & -0.355 & [-0.404, -0.302] & 0/8 & 0.0078 & 0.0234 \\
$q=20$ & 2.012 & 0.678 & 0.616 & 0.511 & 0.383 & 0.547 & -0.141 & [-0.174, -0.103] & 0/8 & 0.0078 & 0.0234 \\
$q=30$ & 2.462 & 0.886 & 0.818 & 0.639 & 0.427 & 0.692 & +0.004 & [-0.019, +0.032] & 4/8 & 0.812 & 0.812 \\
$q=40$ & 2.996 & 0.943 & 0.883 & 0.683 & 0.436 & 0.736 & +0.048 & [+0.023, +0.078] & 7/8 & 0.0156 & 0.0312 \\
$q=50$ & 3.695 & 0.960 & 0.897 & 0.683 & 0.429 & 0.742 & +0.054 & [+0.031, +0.081] & 8/8 & 0.0078 & 0.0234 \\
$q=60$ & 4.702 & 0.965 & 0.889 & 0.666 & 0.418 & 0.734 & +0.046 & [+0.029, +0.067] & 8/8 & 0.0078 & 0.0234 \\
$q=70$ & 6.348 & 0.961 & 0.872 & 0.646 & 0.406 & 0.721 & +0.033 & [+0.022, +0.048] & 8/8 & 0.0078 & 0.0234 \\
$q=80$ & 9.591 & 0.954 & 0.854 & 0.628 & 0.397 & 0.708 & +0.020 & [+0.013, +0.030] & 8/8 & 0.0078 & 0.0234 \\
$q=90$ & 19.236 & 0.946 & 0.841 & 0.615 & 0.389 & 0.698 & +0.010 & [+0.005, +0.018] & 8/8 & 0.0078 & 0.0234 \\
uncapped & -- & 0.942 & 0.827 & 0.602 & 0.381 & 0.688 & reference & & & & \\
\midrule
\multicolumn{12}{l}{\textit{Sampled magnitude}} \\
$q=10$ & 1.618 & 0.291 & 0.243 & 0.227 & 0.204 & 0.241 & -0.437 & [-0.467, -0.409] & 0/8 & 0.0078 & 0.0234 \\
$q=20$ & 2.002 & 0.516 & 0.473 & 0.395 & 0.307 & 0.423 & -0.256 & [-0.288, -0.221] & 0/8 & 0.0078 & 0.0234 \\
$q=30$ & 2.418 & 0.753 & 0.669 & 0.516 & 0.363 & 0.575 & -0.103 & [-0.123, -0.083] & 0/8 & 0.0078 & 0.0234 \\
$q=40$ & 2.919 & 0.870 & 0.772 & 0.578 & 0.385 & 0.651 & -0.027 & [-0.041, -0.013] & 0/8 & 0.0078 & 0.0234 \\
$q=50$ & 3.584 & 0.916 & 0.814 & 0.602 & 0.392 & 0.681 & +0.002 & [-0.007, +0.013] & 5/8 & 0.547 & 0.547 \\
$q=60$ & 4.549 & 0.932 & 0.827 & 0.606 & 0.392 & 0.689 & +0.010 & [+0.004, +0.017] & 7/8 & 0.0234 & 0.0469 \\
$q=70$ & 6.131 & 0.935 & 0.828 & 0.602 & 0.389 & 0.688 & +0.010 & [+0.006, +0.014] & 8/8 & 0.0078 & 0.0234 \\
$q=80$ & 9.265 & 0.935 & 0.825 & 0.597 & 0.385 & 0.685 & +0.007 & [+0.005, +0.009] & 8/8 & 0.0078 & 0.0234 \\
$q=90$ & 18.572 & 0.933 & 0.820 & 0.593 & 0.382 & 0.682 & +0.003 & [+0.002, +0.005] & 8/8 & 0.0078 & 0.0234 \\
uncapped & -- & 0.932 & 0.815 & 0.588 & 0.379 & 0.679 & reference & & & & \\
\bottomrule\end{tabular}\end{table}

\clearpage
\begin{center}\scriptsize\setlength{\tabcolsep}{4pt}
\begin{longtable}{llrcccc}
\caption{\textbf{Comparisons of noise-prune to the other methods at each sparsity level.} Effect size is the mean over tasks of the task-by-task difference in retention at the given sparsity (first method minus second), with its $95\%$ bootstrap confidence interval, the number of tasks favoring the first method, and the exact two-sided sign-flip $p$-value. The corresponding comparisons across the whole pruning range are given in Supplementary Table~\ref{tab:S_main}.}
\label{tab:S_sparsity}\\
\toprule
Method & Compared to & Sparsity & Effect & 95\% CI & Tasks & $p$ \\\midrule
\endfirsthead
\multicolumn{7}{l}{\footnotesize Supplementary Table~\ref{tab:S_sparsity} (continued)}\\\toprule
Method & Compared to & Sparsity & Effect & 95\% CI & Tasks & $p$ \\\midrule
\endhead
\bottomrule\endlastfoot
Noise-prune, capped rescale & Noise-prune, rescale & 10\% & +0.000 & [-0.000, +0.000] & 4/8 & 0.859 \\
 &  & 20\% & +0.000 & [-0.002, +0.002] & 4/8 & 0.773 \\
 &  & 30\% & +0.001 & [-0.002, +0.005] & 5/8 & 0.367 \\
 &  & 40\% & +0.006 & [-0.001, +0.013] & 6/8 & 0.125 \\
 &  & 50\% & +0.020 & [+0.005, +0.035] & 6/8 & 0.0312 \\
 &  & 60\% & +0.059 & [+0.026, +0.102] & 7/8 & 0.0156 \\
 &  & 70\% & +0.083 & [+0.050, +0.115] & 8/8 & 0.0078 \\
 &  & 80\% & +0.039 & [+0.012, +0.062] & 7/8 & 0.0234 \\
 & Noise-prune, score only & 10\% & +0.000 & [-0.003, +0.004] & 5/8 & 0.953 \\
 &  & 20\% & +0.007 & [-0.005, +0.019] & 5/8 & 0.281 \\
 &  & 30\% & +0.027 & [+0.005, +0.054] & 6/8 & 0.0547 \\
 &  & 40\% & +0.043 & [+0.015, +0.083] & 8/8 & 0.0078 \\
 &  & 50\% & +0.086 & [+0.047, +0.126] & 8/8 & 0.0078 \\
 &  & 60\% & +0.305 & [+0.222, +0.394] & 8/8 & 0.0078 \\
 &  & 70\% & +0.382 & [+0.325, +0.436] & 8/8 & 0.0078 \\
 &  & 80\% & +0.273 & [+0.160, +0.414] & 8/8 & 0.0078 \\
 & Sampled magnitude, capped rescale & 10\% & +0.000 & [-0.000, +0.001] & 6/8 & 0.430 \\
 &  & 20\% & -0.000 & [-0.002, +0.002] & 3/8 & 0.781 \\
 &  & 30\% & +0.003 & [-0.001, +0.006] & 6/8 & 0.0859 \\
 &  & 40\% & +0.009 & [-0.003, +0.019] & 5/8 & 0.156 \\
 &  & 50\% & +0.026 & [+0.006, +0.046] & 7/8 & 0.0234 \\
 &  & 60\% & +0.059 & [+0.027, +0.094] & 7/8 & 0.0156 \\
 &  & 70\% & +0.079 & [+0.045, +0.113] & 8/8 & 0.0078 \\
 &  & 80\% & +0.035 & [+0.013, +0.054] & 7/8 & 0.0156 \\
 & Sampled magnitude & 10\% & +0.000 & [-0.000, +0.001] & 5/8 & 0.695 \\
 &  & 20\% & +0.000 & [-0.002, +0.002] & 4/8 & 0.984 \\
 &  & 30\% & +0.003 & [-0.000, +0.007] & 6/8 & 0.0469 \\
 &  & 40\% & +0.010 & [-0.001, +0.021] & 6/8 & 0.109 \\
 &  & 50\% & +0.029 & [+0.009, +0.050] & 7/8 & 0.0234 \\
 &  & 60\% & +0.072 & [+0.035, +0.111] & 7/8 & 0.0156 \\
 &  & 70\% & +0.096 & [+0.055, +0.134] & 8/8 & 0.0078 \\
 &  & 80\% & +0.041 & [+0.009, +0.068] & 6/8 & 0.0469 \\
 & Magnitude & 10\% & -0.000 & [-0.001, +0.000] & 4/8 & 0.742 \\
 &  & 20\% & -0.002 & [-0.004, +0.001] & 4/8 & 0.664 \\
 &  & 30\% & +0.007 & [-0.014, +0.025] & 7/8 & 0.414 \\
 &  & 40\% & +0.054 & [+0.005, +0.096] & 7/8 & 0.0391 \\
 &  & 50\% & +0.213 & [+0.143, +0.281] & 8/8 & 0.0078 \\
 &  & 60\% & +0.470 & [+0.392, +0.541] & 8/8 & 0.0078 \\
 &  & 70\% & +0.496 & [+0.439, +0.555] & 8/8 & 0.0078 \\
 &  & 80\% & +0.332 & [+0.197, +0.498] & 8/8 & 0.0078 \\
 & Recurrent OBS & 10\% & -0.000 & [-0.001, +0.000] & 5/8 & 0.578 \\
 &  & 20\% & -0.001 & [-0.004, +0.001] & 4/8 & 0.219 \\
 &  & 30\% & -0.001 & [-0.009, +0.005] & 4/8 & 0.688 \\
 &  & 40\% & -0.003 & [-0.026, +0.013] & 5/8 & 0.844 \\
 &  & 50\% & +0.031 & [-0.008, +0.060] & 7/8 & 0.0625 \\
 &  & 60\% & +0.162 & [+0.108, +0.210] & 8/8 & 0.0078 \\
 &  & 70\% & +0.303 & [+0.200, +0.391] & 8/8 & 0.0078 \\
 &  & 80\% & +0.282 & [+0.171, +0.416] & 8/8 & 0.0078 \\
 & Random & 10\% & +0.635 & [+0.530, +0.750] & 8/8 & 0.0078 \\
 &  & 20\% & +0.869 & [+0.822, +0.917] & 8/8 & 0.0078 \\
 &  & 30\% & +0.931 & [+0.910, +0.954] & 8/8 & 0.0078 \\
 &  & 40\% & +0.941 & [+0.923, +0.960] & 8/8 & 0.0078 \\
 &  & 50\% & +0.906 & [+0.872, +0.942] & 8/8 & 0.0078 \\
 &  & 60\% & +0.822 & [+0.753, +0.895] & 8/8 & 0.0078 \\
 &  & 70\% & +0.607 & [+0.464, +0.770] & 8/8 & 0.0078 \\
 &  & 80\% & +0.356 & [+0.166, +0.586] & 8/8 & 0.0078 \\
\midrule
Noise-prune, rescale & Noise-prune, score only & 10\% & +0.000 & [-0.004, +0.004] & 5/8 & 0.984 \\
 &  & 20\% & +0.006 & [-0.006, +0.020] & 5/8 & 0.352 \\
 &  & 30\% & +0.025 & [+0.004, +0.052] & 6/8 & 0.0547 \\
 &  & 40\% & +0.037 & [+0.009, +0.077] & 8/8 & 0.0078 \\
 &  & 50\% & +0.067 & [+0.029, +0.105] & 8/8 & 0.0078 \\
 &  & 60\% & +0.245 & [+0.164, +0.329] & 8/8 & 0.0078 \\
 &  & 70\% & +0.299 & [+0.251, +0.351] & 8/8 & 0.0078 \\
 &  & 80\% & +0.234 & [+0.130, +0.374] & 8/8 & 0.0078 \\
 & Sampled magnitude, capped rescale & 10\% & +0.000 & [-0.000, +0.001] & 5/8 & 0.586 \\
 &  & 20\% & -0.000 & [-0.001, +0.001] & 4/8 & 0.656 \\
 &  & 30\% & +0.001 & [-0.004, +0.006] & 4/8 & 0.645 \\
 &  & 40\% & +0.002 & [-0.010, +0.012] & 4/8 & 0.730 \\
 &  & 50\% & +0.006 & [-0.008, +0.019] & 5/8 & 0.453 \\
 &  & 60\% & -0.000 & [-0.027, +0.023] & 4/8 & 0.977 \\
 &  & 70\% & -0.003 & [-0.027, +0.023] & 3/8 & 0.820 \\
 &  & 80\% & -0.004 & [-0.019, +0.011] & 2/8 & 0.531 \\
 & Sampled magnitude & 10\% & +0.000 & [-0.000, +0.000] & 5/8 & 0.625 \\
 &  & 20\% & -0.000 & [-0.001, +0.001] & 4/8 & 0.781 \\
 &  & 30\% & +0.002 & [-0.001, +0.005] & 5/8 & 0.109 \\
 &  & 40\% & +0.004 & [-0.001, +0.009] & 5/8 & 0.188 \\
 &  & 50\% & +0.010 & [+0.000, +0.020] & 7/8 & 0.0156 \\
 &  & 60\% & +0.012 & [-0.000, +0.025] & 6/8 & 0.0547 \\
 &  & 70\% & +0.014 & [-0.012, +0.042] & 5/8 & 0.375 \\
 &  & 80\% & +0.002 & [-0.007, +0.011] & 3/8 & 0.773 \\
 & Magnitude & 10\% & -0.000 & [-0.001, +0.000] & 3/8 & 0.820 \\
 &  & 20\% & -0.002 & [-0.005, +0.001] & 4/8 & 0.422 \\
 &  & 30\% & +0.005 & [-0.016, +0.023] & 6/8 & 0.547 \\
 &  & 40\% & +0.048 & [+0.001, +0.088] & 7/8 & 0.0391 \\
 &  & 50\% & +0.194 & [+0.129, +0.258] & 8/8 & 0.0078 \\
 &  & 60\% & +0.411 & [+0.347, +0.467] & 8/8 & 0.0078 \\
 &  & 70\% & +0.414 & [+0.343, +0.489] & 8/8 & 0.0078 \\
 &  & 80\% & +0.293 & [+0.150, +0.496] & 8/8 & 0.0078 \\
 & Recurrent OBS & 10\% & -0.000 & [-0.001, +0.000] & 3/8 & 0.422 \\
 &  & 20\% & -0.002 & [-0.004, +0.000] & 3/8 & 0.0547 \\
 &  & 30\% & -0.003 & [-0.010, +0.003] & 3/8 & 0.344 \\
 &  & 40\% & -0.010 & [-0.030, +0.006] & 3/8 & 0.359 \\
 &  & 50\% & +0.012 & [-0.023, +0.041] & 5/8 & 0.578 \\
 &  & 60\% & +0.102 & [+0.039, +0.160] & 7/8 & 0.0312 \\
 &  & 70\% & +0.220 & [+0.105, +0.321] & 8/8 & 0.0078 \\
 &  & 80\% & +0.243 & [+0.137, +0.381] & 8/8 & 0.0078 \\
 & Random & 10\% & +0.635 & [+0.530, +0.750] & 8/8 & 0.0078 \\
 &  & 20\% & +0.869 & [+0.822, +0.917] & 8/8 & 0.0078 \\
 &  & 30\% & +0.929 & [+0.908, +0.952] & 8/8 & 0.0078 \\
 &  & 40\% & +0.935 & [+0.916, +0.956] & 8/8 & 0.0078 \\
 &  & 50\% & +0.886 & [+0.845, +0.927] & 8/8 & 0.0078 \\
 &  & 60\% & +0.763 & [+0.687, +0.846] & 8/8 & 0.0078 \\
 &  & 70\% & +0.525 & [+0.374, +0.697] & 8/8 & 0.0078 \\
 &  & 80\% & +0.317 & [+0.138, +0.567] & 8/8 & 0.0078 \\
\end{longtable}\end{center}